\documentclass[utf8]{frontiersFPHY} 

\usepackage{url,hyperref,lineno,microtype,subcaption}
\usepackage[onehalfspacing]{setspace}

\usepackage{mystylefile} 

\usepackage{xr}	
\makeatletter
\newcommand*{\addFileDependency}[1]{
	\typeout{(#1)}
	\@addtofilelist{#1}
	\IfFileExists{#1}{}{\typeout{No file #1.}}
}
\makeatother

\def\keyFont{\fontsize{8}{11}\helveticabold }
\def\firstAuthorLast{Chi {et~al.}} 
\def\Authors{Jocelyn T. Chi\,$^{1}$, Ilse C. F. Ipsen\,$^{2}$, 
Tzu-Hung Hsiao\,$^{3}$,
Ching-Heng Lin\,$^{3}$,
Li-San Wang\,$^{4}$,
Wan-Ping Lee\,$^{4}$,
Tzu-Pin Lu\,$^{5}$,
and Jung-Ying Tzeng\,$^{1,5,6*}$}
\def\Address{$^{1}$Department of Statistics, North Carolina State University, Raleigh, NC, USA \\
$^{2}$Department of Mathematics, North Carolina State University, Raleigh, NC, USA\\
$^{3}$Department of Medical Research, Taichung Veterans General Hospital, Taichung, Taiwan\\
$^{4}$
Penn Neurodegeneration Genomics Center, Department of Pathology and Laboratory Medicine, Perelman School of Medicine, University of Pennsylvania, Philadelphia, PA, USA\\
$^{5}$Institute of Epidemiology and Preventive Medicine, National Taiwan University, Taipei, Taiwan\\
$^{6}$Department of Statistics, National Cheng-Kung University, Tainan, Taiwan
}
\def\corrAuthor{Jung-Ying Tzeng}

\def\corrEmail{jytzeng@ncsu.edu}

\begin{document}
\onecolumn
\firstpage{1}

\title[SEAGLE: A Scalable Exact Algorithm for Large-Scale Set-Based \GxE Tests in Biobank Data]{SEAGLE: A Scalable Exact Algorithm for Large-Scale Set-Based Gene-Environment Interaction Tests in Biobank Data} 

\author[\firstAuthorLast ]{\Authors} 
\address{} 
\correspondance{} 

\extraAuth{}

\maketitle

\begin{abstract}
The explosion of biobank data offers unprecedented opportunities for gene-environment interaction (GxE) studies of complex diseases because of the large sample sizes and the rich collection in genetic and non-genetic information. However, the extremely large sample size also introduces new computational challenges in \GxE assessment, especially for set-based \GxE variance component (VC) tests, which are a widely used strategy to boost overall \GxE signals and to evaluate the joint \GxE effect of multiple variants from a biologically meaningful unit (e.g., gene). In this work, we focus on continuous traits and present SEAGLE, a \textbf{S}calable \textbf{E}xact \textbf{A}l\textbf{G}orithm for \textbf{L}arge-scale set-based Gx\textbf{E} tests, to permit \GxE VC tests for biobank-scale data. 
SEAGLE employs modern matrix computations to calculate the test statistic and p-value of the GxE VC test in a computationally efficient fashion, 
without imposing additional assumptions or relying on approximations. SEAGLE can easily accommodate sample sizes in the order of $10^5$, is implementable on standard laptops, and does not require specialized computing equipment. We demonstrate the performance of SEAGLE using extensive simulations.  We illustrate its utility by conducting genome-wide gene-based \GxE analysis on the Taiwan Biobank data to explore the interaction of gene and physical activity status on body mass index. 
%

\tiny
 \keyFont{ \section{Keywords:} scalable GEI test, gene-environment variance component test,  gene-environment kernel test, regional-based gene-environment test, gene-based GxE test for biobank data, GxE collapsing test for biobank data, GxE test for large-scale sequencing data} 
\end{abstract}

\section{Introduction}
\label{sec:intro}

Human complex diseases such as neurodegenerative diseases, psychiatric disorders, metabolic syndromes, and cancers are complex traits for which disease susceptibility, disease development, and treatment response are mediated by intricate genetic and environmental factors. Understanding the genetic etiology of these complex diseases requires collective consideration of potential genetic and environmental contributors. Studies of gene-environment interactions (G$\times$E) enable understanding of the differences that environmental exposures may have on health outcomes in people with varying genotypes \citep{ottman1996gene, hunter2005gene, mcallister2017current}. Examples include
the impact of physical activity and alcohol consumption on the genetic risk for obesity-related traits~\citep{sulc2020quantification}, the impact of
air pollution on the genetic risk for cardio-metabolic and respiratory traits~\citep{fave2018gene}, and other examples  reviewed in \cite{ritz2017lessons_PMID:28978190}. 

%

When assessing \GxE effects, set-based tests are  popular approaches to detecting interactions between an environmental factor and  a set of single nucleotide polymorphism (SNPs)
in a gene, sliding window, or functional region~\citep{lin2013testGESAT,  su2017_PMID:27474101,lin2016test, tzeng2011studying,zhao2015assessing}. Compared to single-SNP \GxE tests, set-based \GxE tests can enhance testing performance by reducing multiple-testing burden and by aggregating \GxE signals over multiple SNPs that are of moderate effect sizes or of low frequencies.

Large-scale biobanks collect genetic and health information on hundreds of thousands of individuals. Their large sample sizes and rich data on non-genetic factors offer unprecedented opportunities for in-depth studies on \GxE effects.  While the explosion of biobank data collections provides great hopes for novel \GxE discoveries, it also introduces computational challenges.
In particular, many set-based \GxE tests can be cast as variance component (VC) tests 
under a random effects modeling framework \citep{lin2013testGESAT, su2017_PMID:27474101}, including kernel machine based tests~\citep{lin2016test, wang2015_PMID:25538034} and  similarity regression based methods~\citep{tzeng2011studying, zhao2015assessing}.
Hypothesis testing in this framework relies on computations 
with phenotypic variance matrices with dimension $n \times n$ (with $n$ as the sample size) and may involve estimating nuisance variance components. 
When $n$ is large, as in the case of biobank data, matrix computations whose operation counts scale with $n^3$ are prohibitive in terms of computation time and storage.

A number of methods attempt to ease this computational burden by bypassing the estimation of nuisance variance components, either through approximation of the variance or kernel matrices~\citep{marceau2015fast} or through approximation of the score-like test statistics~\citep{wang2020efficient}.  In the first case, approximating the kernel matrices still 
requires an expensive eigenvalue decomposition upfront, in addition to storage 
for the explicit formation of the $n \times n$ kernel matrices, thus lacking practical scalability.
In the latter case, approximating the test statistics requires assumptions that may or may not 
be valid and are difficult to validate in practice. Our numerical studies in Section~\ref{sec:numexp} show that the Type 1 error rates and power can be sub-optimal when data do not adhere to the required assumptions.

In this work, we focus on continuous traits and introduce a {\bf S}calable {\bf E}xact {\bf A}l{\bf G}orithm for {\bf L}arge-scale set-based G$\times${\bf E} tests (SEAGLE) for performing \GxE VC tests on biobank data.
Here ``exact" refers to the fact that SEAGLE computes the original VC test statistic without any approximations, rather than the null distribution 
of the test statistic being asymptotic or exact. 
Exactness and scalability are achieved through the judicious use of 
modern matrix computations, allowing us to dispense with approximations and assumptions.  
Our numerical experiments illustrate that SEAGLE 
produces Type 1 error rates and power identical to those of the original GxE VC methods~\citep{tzeng2011studying}, but at a fraction of the speed.  Additionally, SEAGLE can easily handle biobank-scale data with as many as 
$n$ individuals in the order of $10^5$, 
often at the same speed as state-of-the-art approximate methods~\citep{wang2020efficient}.
Compared with the state-of-the-art approximate method~\citep{wang2020efficient},
SEAGLE can produce more accurate  Type 1 error rates and power.
Another advantage of SEAGLE is its user-friendliness; it can be run on ordinary laptops and does not require specialized or high performance computing equipment or parallelization.  In fact, our timing comparisons in Section \ref{sec:numexp} were performed on a 2013 Intel Core i5 laptop with a 2.70 GHz CPU and 16 GB RAM, specs that are standard for modern laptops.  Therefore, SEAGLE 
makes it possible to run exact and scalable \GxE VC tests on biobank-scale data
with just a modicum of computational resources. 


The rest of the paper proceeds as follows.  Section \ref{sec:method} describes the standard mixed effects model for \GxE effects, testing procedures, computational performance, and SEAGLE algorithm. Section~\ref{sec:numexp} illustrates SEAGLE's performance through numerical studies.  Section \ref{sec:discussion} concludes with a brief 
 summary of our contributions and avenues for future work.

\section{Materials and Methods}
\label{sec:method}
We describe the standard mixed effects model for \GxE effects and testing procedures (Section~\ref{sec:test}), 
the computational challenges for biobank-scale data (Section~\ref{sec:challenges}), 
the components of the SEAGLE algorithm (Section \ref{sec:alg}), and the SEAGLE
algorithm as a whole (Section~\ref{sec:together}).
\bigskip

\subsection{GxE Variance Component Tests for Continuous Traits} 
\label{sec:test}
We present the standard mixed effects model for studying \GxE effects, the score-like test statistic,
and its p-value.
Let $\vy \in \mathbb{R}^{n}$ denote the response vector with $n$ individual responses for a 
continuous trait;  $\mx \in \mathbb{R}^{n \times p}$ the design matrix of $p$ covariates
whose leading column  is the all ones vector for the intercept;
$\me \in \mathbb{R}^{n}$ the design vector of the environmental factor in the \GxE effect;
and $\mg \in \mathbb{R}^{n \times L}$ the genetic marker matrix for the $L$ SNPs where $L< n$. Define
the design matrix for the \GxE terms as
$\mtg = \diag(\me)\mg\in\mathbb{R}^{n \times L}$ where $\diag(\me)\in\mathbb{R}^{n\times n}$ 
is a diagonal matrix with the elements of the vector~$\me$ on the diagonal.

Consider the linear mixed effects model \citep{tzeng2011studying,lin2013testGESAT},
\begin{eqnarray}
	\label{eqn:km}
	\vy = \mx\vbeta_{X} + \me\vbeta_{E} + \mg \vb + \mtg \vc + \veps.
\label{eqn:mixedmodel}
\end{eqnarray}   
Here,
$\vbeta_{X}\in \mathbb{R}^{p}$ and $\vbeta_{E}\in \mathbb{R}$ are the fixed-effects coefficients for the covariates and environmental factor, respectively;  $\vb \in \mathbb{R}^{L}$ and $\vc \in \mathbb{R}^{L}$ are the genetic main (G) effect and \GxE effect, respectively, with $\vb \sim \text{N}(\vzero, \tau \mi_L)$ and $\vc \sim \text{N}(\vzero, \nu \mi_L)$;  $\veps \sim \text{N}(\vzero, \sigma\, \mi_{n})$; and $\mi_k\in\mathbb{R}^{k\times k}$ denotes the identity matrix
of dimension $k$.

The SNP-set analysis models the G and \GxE effects of the $L$ SNPs as random effects rather than fixed
effects. This choice avoids power loss for non-small $L$
and numerical difficulties from correlated SNPs that can occur in a fixed effects model. To assess the presence 
of \GxE effects with $H_0:\vc={\bf 0}$ in Model~\eqref{eqn:mixedmodel},
one can apply a score-like test to the corresponding variance component with   $H_{0}: \nu = 0$.

To simplify  the null model of \eqref{eqn:mixedmodel} in the score-like test, we consolidate and define $\mtx = \begin{pmatrix} \mx & \me \end{pmatrix} \in \mathbb{R}^{n \times P}$ 
and $\vbeta = \begin{pmatrix} \vbeta_{\mx}\Tra & \vbeta_{\me}\Tra \end{pmatrix}\Tra \in\mathbb{R}^P$, where $P=p+1$.
The resulting null model becomes $\vy = \mtx\vbeta + \mg\vb + \veps$, where the response is
$\vy \sim \text{N}\left( \mtx\vbeta \mycomma \mv \right)$ with  $\mv = \tau\,\mg \mg\Tra + \sigma\,\mi_{n}$.
Following \citep{tzeng2011studying}, the score-like test statistic is 
\begin{eqnarray}\label{eqn:scoretest}
	T &=&
	\frac{1}{2}\, (\vy - \vhmu)\Tra \mv\Inv \mtg \mtg\Tra \mv\Inv (\vy - \vhmu)\nonumber\\
	&=&\frac{1}{2}\, \vy\Tra \mP  \mtg \mtg\Tra \mP \vy
	\equiv \frac{1}{2}\vt\Tra\vt, 
	\text{\ \   where\ \   } \vt = \mtg\Tra\mP\vy.
\end{eqnarray}
In Equation~\eqref{eqn:scoretest}, 
$\vhmu = \mtx\vhbeta =\mtx(\mtx\Tra\mv\Inv\mtx)\Inv\mtx\Tra\mv\Inv\vy$ and
$\mP = \mv\Inv - \mv\Inv \mtx(\mtx\Tra\mv\Inv\mtx)\Inv\mtx\Tra\mv\Inv$. 
Appendix \ref{app:em-reml} presents
the restricted maximum likelihood (REML) expectation-maximization (EM) algorithm for estimating the nuisance VC parameters $\tau$ and $\sigma$ for computing  $T$~\citep{tzeng2011studying, zhao2015assessing}.
The test statistic $T$ follows a weighted $\chi^2_{(1)}$ distribution asymptotically under $H_0:\nu=0$.
That is, $T\sim \sum_{\ell} \lambda_{\ell}\chi^2_{(1)}$, where $\lambda_{\ell}$'s are the 
  eigenvalues of 
  \begin{equation}\label{e_evc}
\mc 
=\mc_1\mc_1^T, \qquad \text{where} \quad \mc_1=\frac{1}{\sqrt{2}}\mv^{\frac{1}{2}}\mP\tilde{\mg}.
\end{equation}
Given the $\lambda_{\ell}$'s, 
the p-value of $T$ can be computed with the moment matching method in~\cite{liu2009anew} or 
the exact method in~\cite{davis1980}.
\bigskip

\subsection{Computational Challenges in GxE VC Tests for Biobank-Scale Data}
\label{sec:challenges}
We identify three computational bottlenecks.
\begin{enumerate}
\item The test statistic $T$ 
and the p-value computation
depend on $\mP \in \mathbb{R}^{n \times n}$, which in turn depends on 
$\mv\Inv \in \mathbb{R}^{n \times n}$.  
Explicit formation of the inverse is too expensive and numerically inadvisable, due to loss of
numerical accuracy and stability \cite[Chapter 14]{Higham2002}.

\item The REML EM algorithm (Appendix~\ref{app:em-reml}) estimates the nuisance variance 
components $\tau$ and $\sigma$
in $\mv$ under the null hypothesis. Each iteration requires products with the orthogonal
projector $\mi - \mtx(\mtx\Tra\mtx)\Inv\mtx\Tra$, and inverting a matrix of 
dimension $n-P\approx n$.

\item Computing the p-values requires two eigenvalue decompositions:
(1)  an eigenvalue decomposition of $\mv$ to compute $\mv^{\frac{1}{2}}$
in $\mc_{1}$
; and (2) an eigenvalue decomposition of $\mc$ to 
compute the $\lambda_{\ell}$'s in the weighted $\chi^2_{(1)}$ distribution. 
Computing the eigenvalues and eigenvectors  of the symmetric matrix
$\mv\in\mathbb{R}^{n\times n}$ requires 
$\mathcal{O}(n^{3})$ arithmetic operations and $\mathcal{O}(n^3)$ storage. 
Computing the eigenvalues of $\mc\in\mathbb{R}^{n\times n}$  
requires another $\mathcal{O}(n^3)$ arithmetic operations.
\end{enumerate}
\bigskip

\subsection{Components of the SEAGLE Algorithm for Biobank-Scale GxE VC Test}
\label{sec:alg}
We present our approach for overcoming the three computational challenges in the previous section:
Multiplication with $\mv^{-1}$ without explicit formation of the inverse (Section~\ref{sec:alg-vinv}),
a scalable REML EM algorithm (Section~\ref{sec:REML}),
and a scalable algorithm for computing the eigenvalues of $\mc$ (Section~\ref{sec:evC}).
The idea is to replace explicit formation 
of inverses by low-rank updates and linear system solutions; and to replace $n \times n$ eigenvalue 
decompositions with $L \times L$ ones.
\bigskip

\subsubsection{Multiplication by $\mv\Inv$ without explicit formation of $\mv\Inv$}
\label{sec:alg-vinv}
The test statistic $T$ 
and its p-value calculation depend on $\mv\Inv$. We avoid the explicit formation of the inverse by 
viewing $\mv = \tau\,\mg \mg\Tra + \sigma\,\mi_{n}$
as the low-rank update of a diagonal matrix, and then applying the
Sherman-Morrison-Woodbury formula below  to reduce the
dimension of the computed inverse from $n$ to $L$ where $L\ll n$.

\begin{lemma}[Section 2.1.4 in \cite{golub2013matrix}]\label{lem:woodbury}
	Let $\mh \in \mathbb{R}^{n \times n}$ be nonsingular, and 
	let $\mU, \mb \in \mathbb{R}^{n \times L}$ so that $\mi + \mb\Tra \mh\Inv \mU$ is nonsingular.
	Then
	$$
	(\mh + \mU\mb\Tra)\Inv = \mh\Inv - \mh\Inv\mU(\mi + \mb\Tra \mh\Inv \mU)\Inv \mb\Tra \mh\Inv.
	$$
\end{lemma}

Applying Lemma~\ref{lem:woodbury} to the product of the inverse of
	$\mv = \sigma\left(\mi_{n} + \frac{\tau}{\sigma}\,\mg\mg\Tra\right)$
with any right-hand side input $\mw \in \mathbb{R}^{n \times l}$ gives
\begin{eqnarray}
	\mv\Inv \mw &=& \frac{1}{\sigma}\left[\mw - \frac{\tau}{\sigma}\,\mg\left(\mi_{L} + \frac{\tau}{\sigma}\,\mg\Tra\mg\right)\Inv\mg\Tra\mw\right],
	\label{eqn:alg1}
\end{eqnarray}
which reduces the dimension of the inverse from $n$ to $L$. The explicit computation 
of the inverse of $\mm = \mi_{L} + \frac{\tau}{\sigma}\,\mg\Tra\mg\in\mathbb{R}^{L\times L}$ is, 
in turn, avoided
with a Cholesky decomposition followed by a linear system solution.
Algorithm \ref{alg:applyVinv} shows pseudocode for computing~\eqref{eqn:alg1}.
As a further saving, we pre-compute the Cholesky factorization of $\mm$ only once, 
so it is available for re-use in the computation of the test statistic and p-value. 

\begin{algorithm}[ht]
	\caption{\texttt{applyVinv}} \label{alg:applyVinv}
	
	{\bf Input:} $\mg \in \mathbb{R}^{n \times L}$,\; $\mw \in \mathbb{R}^{n\times l}$,\; $\hat{\tau} > 0$,\;
	$\hat{\sigma} > 0$  \\
	{\bf Output:} $\mv\Inv\mw$\\
	
	\begin{algorithmic}
		\STATE $\mm = \mi_{L} + \frac{\tau}{\sigma} \mg\Tra \mg$
		\STATE Cholesky decomposition $\mm = \ml \ml\Tra$  \COMMENT{$\ml$ is lower triangular}
		\STATE Solve $\ml \mx_{1} = \mg^T \mw$ \COMMENT{Solve lower triangular system for $\mx_1$}
		\STATE Solve $\ml^T \mx_{2} = \mx_{1}$ \COMMENT{Solve upper triangular system for $\mx_2$}
		\STATE \COMMENT{$\qquad$ to obtain $\mx_{2}=\mm\Inv \mg\Tra \mw$}
		\RETURN $\mx = \frac{1}{\sigma}(\mw - \frac{\tau}{\sigma} \mg\mx_{2})$
		\COMMENT{Return $\mv\Inv\mw$  in~\eqref{eqn:alg1}}
	\end{algorithmic}
\end{algorithm}
\bigskip


\subsubsection{Scalable REML EM Algorithm}\label{sec:REML}
We present the scalable version of the REML EM algorithm in Appendix \ref{app:em-reml}
that avoids explicit formation of the orthogonal projector and the inverses.

We assume throughout that $\mtx$ has full column rank with $\rank(\mtx) = P$, and let
$\range(\mtx)$ be the space spanned by the columns of $\mtx$. The space
perpendicular to $\range(\mtx)$ is $\range(\mtx)^{\perp}$, and the orthogonal projector onto this space 
is 
\begin{equation}\label{eq_a}
\ma\ma\Tra = \mi - \mtx(\mtx\Tra\mtx)\Inv\mtx\Tra
\end{equation}
where $\ma\in\mathbb{R}^{n\times (n-P)}$ has orthonormal columns with $\ma\Tra\ma = \mi_{n-P}$.
Let $\vu = \ma\Tra\vy\in\mathbb{R}^{n-P}$ be the orthogonal projection of the response onto
$\range(\mtx)^{\perp}$.

In iteration $t+1$ of the algorithm,  define 
\begin{equation}\label{eq_r}
\mhr = \tauhatt\, \ma\Tra\mg\mg\Tra\ma + \sigmahatt\, \mi_{n-P} \in\mathbb{R}^{(n-P)\times (n-P)}.
\end{equation}
Then the updates in \eqref{eqn:tauupdate} and~\eqref{eqn:sigmaupdate} from Appendix~\ref{app:em-reml} 
can be expressed as 
\begin{eqnarray*}
	\tauhattp &=& \frac{\tauhatt}{L}\left[ \tauhatt\, \|\mg\Tra\ma\mhr\Inv\vu\|^{2}_{2} + \trace(\mi_{L}-\tauhatt\mg\Tra\ma\mhr\Inv\ma\Tra\mg) \right]\\
	\sigmahattp &=& \frac{\sigmahatt}{n-P} \left[ \|\mhr\Inv\vu\|^{2}_{2} + \tauhatt\,\trace(\mg\Tra\ma\mhr\Inv\ma\Tra\mg) \right].
\end{eqnarray*}
The two bottlenecks in the REML EM algorithm are the computation of the non-symmetric 
``square-root"~$\ma$ in (\ref{eq_a}), and products with $\mhr^{-1}$ from (\ref{eq_r}).
Since $P$ is small, $n-P\approx n$, hence explicit formation of the
inverse is out of the question, especially since $\mhr$ changes in each iteration
due to the updates for $\tauhatt$ and $\sigmahatt$.

To avoid explicit formation of the full matrix in (\ref{eq_a}),
we compute instead the QR decomposition 
\begin{eqnarray}\label{e_qr}
	\mtx &=& \underbrace{\begin{pmatrix} \mq_{1} & \mq_{2} \end{pmatrix}}_{\mq}
	\begin{pmatrix}\mr_{0} \\ \vzero\end{pmatrix},
\end{eqnarray}
where $\mq\in\mathbb{R}^{n\times n}$ is an orthogonal matrix with $\mq^T\mq=\mq\mq^T=\mi_n$.
The columns of $\mq_{1} \in \mathbb{R}^{n \times P}$ form an orthonormal basis for $\range(\mtx)$, 
and the columns of $\mq_{2} \in \mathbb{R}^{n \times (n-P)}$ form an orthonormal basis for $\range(\mtx)^{\perp}$. The upper triangular matrix 
$\mr_0\in\mathbb{R}^{P\times P}$ is nonsingular, due to the assumption of $\mtx$ having
full column rank. Therefore, (\ref{eq_a}) simplifies to
\begin{eqnarray*}
\mi - \mtx(\mtx\Tra\mtx)\Inv\mtx\Tra=\mi-\mq_1\mq_1^T=\mq_2\mq_2^T.
\end{eqnarray*}
Thus,  $\ma=\mq_2$ represents the trailing $n-P$ columns 
of the orthogonal matrix $\mq$
in the QR factorization of $\mtx$.

%

Algorithm \ref{alg:REML-EM} shows pseudocode for the REML EM algorithm.
Since $\ma=\mq_2$ occurs only as $\ma^T$ in matrix-vector
or matrix-matrix multiplications,
we do not compute $\ma$ explicitly.
Instead, we compute the full QR decomposition 
in~(\ref{e_qr}) where the ``QR object"
$\mq=\begin{pmatrix} \mq_1 & \ma\end{pmatrix}$ is stored implicitly in factored form. To compute 
$\vu = \ma^T\vy$, we multiply 
$\tilde{\vu}=\mq^T\vy=\left({\mq_1^T\vy\atop \ma^T\vy}\right)$
and then extract the trailing $n-P$ rows from $\tilde{\vu}$.

 Furthermore, we apply $\mhr\Inv$ from (\ref{eq_r}) with a modified version of Algorithm~\ref{alg:applyVinv}
where $\mg$ is replaced by $\ma\Tra\mg$ and $\mi_{n}$ by~$\mi_{n-P}$.  Unfortunately,
one cannot pre-compute a Cholesky factorization for the whole algorithm
since $\tauhatt$ and $\sigmahatt$ change in each iteration. 
However, within a single iteration, we pre-compute a Cholesky factorization of $\mhr$ for
subsequent linear system solutions of $\mhr^{-1}\vu$ and $\mhr^{-1}\ma^T\mg$. Following previous work on \GxE VC tests in \cite{tzeng2011studying}, our convergence 
criteria are: (i) the magnitude of the relative difference between the current and previous estimate; and (ii) the default convergence tolerance from the \texttt{SIMreg} package for R.
\bigskip

\begin{algorithm}[ht]
	\caption{\texttt{REML-EM}} \label{alg:REML-EM}
	
	{\bf Input:} $\vy\in \mathbb{R}^{n}$,\; $\mg \in \mathbb{R}^{n \times L}$,\; $\mtx \in \mathbb{R}^{n \times P}$\; $\hat{\tau}_{0} > 0$,\; $\hat{\sigma}_{0} > 0$  \\
	{\bf Output:} $\hat{\tau}, \hat{\sigma}$\\
	
	\begin{algorithmic}
		\STATE QR decomposition $\mtx = \mq\left({\mr_0\atop\vzero}\right)$
		\COMMENT{Compute QR object in QR decomposition of $\mtx$}
\STATE Multiply $\tilde{\vu}=\mq^T\vy$
		 \COMMENT{Apply \texttt{qr.qty} function to multiply by $\mq\Tra$}
		 \STATE $\vu = \ma\Tra \vy$ 
		 \COMMENT{Extract trailing $n-P$ rows from $\mq^T\vy$}
		\STATE Pre-compute $\ma\Tra \mg$
		\STATE
		\STATE $t=0$
		\WHILE{not converged}
		\STATE Apply $\mhr\Inv$ with modified Algorithm~\ref{alg:applyVinv} and $\tauhatt$ and $\sigmahatt$
		\STATE $\tauhattp = \frac{\tauhatt}{L}\left[ \tauhatt\, \|\mg\Tra\ma\mhr\Inv\vu\|^{2}_{2} + \trace(\mi_{L}-\tauhatt\mg\Tra\ma\mhr\Inv\ma\Tra\mg) \right]$
		\STATE $\sigmahattp = \frac{\sigmahatt}{n-P} \left[ \|\mhr\Inv\vu\|^{2}_{2} + \tauhatt\,\trace(\mg\Tra\ma\mhr\Inv\ma\Tra\mg) \right]$
		\STATE $t = t + 1$
		\ENDWHILE
		\RETURN $\hat{\tau} = \tauhattp$, $\hat{\sigma} = \sigmahattp$
	\end{algorithmic}
\end{algorithm}

\subsubsection{Scalable Algorithm for Computing the Eigenvalues of {\bf C}}
\label{sec:evC}
Computation of the p-values 
requires the eigenvalues of $\mc=\mc_1\mc_1^T$ 
in~(\ref{e_evc}), which in turn 
involves products with $\mv^{\frac{1}{2}}\in \mathbb{R}^{n \times n}$.
We avoid  the computation of the square root by exploiting the fact that the nonzero eigenvalues of $\mc_{1}\mc_{1}\Tra$ are equal to the nonzero eigenvalues of $\mc_{1}\Tra\mc_{1}$. The symmetry of $\mv$ and $\mP$ and the equality $\mP\mv\mP = \mP$ imply the much simpler
expression
	\begin{eqnarray*}
		\mc_{1}\Tra\mc_{1} =\frac{1}{2}\mtg\Tra\mP\mv\mP \mtg 
		=\frac{1}{2}\mtg^T\mP\mtg.
	\end{eqnarray*}
The explicit formation of $\mP$ 
is avoided by computing instead 
products $\mP\mtg$ with Algorithm~\ref{alg:applyVinv}.
Therefore,
our approach of replacing the $n\times n$ matrix 
$\mc_{1}\mc_{1}^T$ with the much smaller $L\times L$ matrix
$\mc_{1}^T\mc_1$ reduces the operation count from
$\mathcal{O}(n^{3})$ down to  $\mathcal{O}(L^{3})$. Part~III of Algorithm \ref{alg:SEAGLE}
shows the pseudocode.
	
\bigskip

\subsubsection{The SEAGLE Algorithm}\label{sec:together}
Combining the algorithms from
Sections \ref{sec:alg-vinv}--\ref{sec:evC}
gives the SEAGLE Algorithm~\ref{alg:SEAGLE}
for computing the score-like 
test statistic $T$ and its p-value. SEAGLE
is implemented in the publicly available R package \texttt{SEAGLE}.

Algorithm~\ref{alg:SEAGLE} consists of three parts.
Part~I computes $\hat{\tau}$ and $\hat{\sigma}$ with
the scalable REML EM in Algorithm~\ref{alg:REML-EM};
Part~II computes the score-like $T$ statistic
in (\ref{eqn:scoretest}); and
Part~III computes the $p$-values
from the eigenvalues of $\mc$ in (\ref{e_evc}).
Linear systems with $\mv$ are efficiently solved with
Algorithm~\ref{alg:applyVinv}.
The fast diagonal multiplication in R 
stores diagonal matrices as vectors.
The QR decomposition is implemented
with the \texttt{qr} function in the R \texttt{base} package. The \texttt{qr.qty} function 
makes it possible to left multiply by $\mq\Tra$ 
without having to explicitly form $\mq$.

\begin{algorithm}
	\caption{\texttt{SEAGLE}} \label{alg:SEAGLE}
	{\bf Input:} $\vy \in \mathbb{R}^{n},\; \mtx\in \mathbb{R}^{n \times P},\; \me \in \mathbb{R}^{n},\; \mg \in \mathbb{R}^{n \times L},\; \hat{\tau} > 0,\; \hat{\sigma} > 0$ \\
	{\bf Output:} $T$,\; p-value\\
	
	Part I: Compute $\hat{\tau}$ and $\hat{\sigma}$ with the scalable REML EM algorithm
	\begin{algorithmic}
		\STATE $\hat{\tau}, \hat{\sigma} = \texttt{REML-EM}(\vy,\; \mg,\; \mtx,\; \hat{\tau}_{0} > 0,\;\hat{\sigma}_{0} > 0)$ \COMMENT{Algorithm~\ref{alg:REML-EM} computes estimates for $\tau$ and $\sigma$
	}
	\end{algorithmic}
	
	Part II: Compute score-like test statistic $T$ in (\ref{eqn:scoretest})
	\begin{algorithmic}
		\STATE $\vx = \texttt{applyVinv}(\mg, \hat{\tau}, \hat{\sigma}, \vy)$
		\COMMENT{Algorithm~\ref{alg:applyVinv} computes $\vx=\mv^{-1}\vy$}
		\STATE $\mh = \texttt{applyVinv}(\mg, \hat{\tau}, \hat{\sigma}, \mtx)$
		\COMMENT{Algorithm~\ref{alg:applyVinv} computes $\mh=\mv^{-1}\mtx$}
		\STATE Multiply $\mgamma = \mtx\Tra\mh$ \COMMENT{$\mgamma=\mtx\Tra\mv\Inv\mtx$} 
		\STATE Cholesky decomposition $\mgamma=\ml\ml^T$ \COMMENT{$\ml$ is lower triangular}
		\STATE Solve $\ml \vc_{1} = \mtx\Tra \vx$ \COMMENT{Solve lower triangular system for $\vc_1$}
		\STATE Solve $\ml^T \vc_{2} = \vc_{1}$ \COMMENT{Solve upper triangular system for $\vc_2$} 
		\STATE  \COMMENT{$\quad$ to obtain $\vc_2=(\mtx\Tra\mv\Inv\mtx)^{-1}\mtx\Tra\mv\Inv\vy$} 
		\STATE $\vy_{\mP} = \texttt{applyVinv}(\mg, \hat{\tau}, \hat{\sigma}, \vy) -
		\texttt{applyVinv}(\mg, \hat{\tau}, \hat{\sigma}, \mtx \vc_{2})$
		\STATE\COMMENT{Compute  $\mP\vy$}
        \STATE $\mtg = \diag(\me)\mg$
		\STATE Multiply $\vt = \mtg\Tra\vy_{\mP}$

		\RETURN $T = \frac{1}{2}\vt\Tra\vt$
	\end{algorithmic}
	
	Part III: Compute p-value from the eigenvalues of $\mc$ in (\ref{e_evc})
	\begin{algorithmic}
		\STATE $\mlambda = \mtx\Tra  \texttt{applyVinv}(\mg, \hat{\tau}, \hat{\sigma}, \mtg)$
		 \COMMENT{$\mlambda=\mtx\Tra\mv\Inv\mtg$}
		\STATE $\ml\mpsi_{1} = \mlambda$ \COMMENT{Solve the lower triangular system for $\mpsi_{1}$}
		\STATE $\ml\Tra\mpsi_{2} = \mpsi_{1}$ \COMMENT{Solve the upper triangular system for $\mpsi_{2}$, and }
		\COMMENT{$\quad$ obtain $\mpsi_{2}=\mgamma\Inv \mlambda$}
		\STATE $\mgamma_{1} = \texttt{applyVinv}(\mg, \hat{\tau}, \hat{\sigma}, \mtg) - \texttt{applyVinv}(\mg, \hat{\tau}, \hat{\sigma}, \mtx \mpsi_{2})$ \\ \COMMENT{Compute product $\mgamma_{1} = \mP\mtg$}
		\STATE $\mgamma_{2} = \frac{1}{2}\,\mtg\mgamma_{1}$
		\COMMENT{Form $\mgamma_{2} = \mc_{1}\Tra\mc_{1}$}
		\STATE Compute eigenvalues $\lambda_{\ell}$'s  of $\mgamma_{2}$ \COMMENT{Compute eigenvalues of $\mc$}
		\RETURN p-value computed from $T$ and $\lambda_{\ell}$'s according to
		\cite{liu2009anew} or \cite{davis1980}.
	\end{algorithmic}
\end{algorithm}

\section{Results}
\label{sec:numexp}
\subsection{Simulation Study}

We evaluate the performance of our proposed method SEAGLE using simulation studies from two settings: I) data simulated from a  random effects genetic model with $n=5{,}000$ observations, and II) data simulated from a fixed effects genetic model with $n=20{,}000$ and $n=100{,}000$ observations.  
In the random effects simulations, we generate data according to Model~\eqref{eqn:mixedmodel}.  This enables us to evaluate SEAGLE's estimation and testing performance.  We consider a smaller $n$ to enable comparisons with existing \GxE VC tests.
In the fixed effects simulations, we generate data from a fixed effects model.  This enables us to evaluate the testing performance when the data do not follow our modeling assumptions.  We consider larger $n$ values to demonstrate SEAGLE's effectiveness on biobank-scale data.

In each setting, we study the Type 1 error rate and power. We consider three baseline approaches: (i) the original \GxE VC test (referred to as OVC)~\citep{tzeng2011studying,wang2014complete}, as implemented in the \texttt{SIMreg} R package (\href{https://www4.stat.ncsu.edu/~jytzeng/software_simreg.php}{\texttt{https://www4.stat.ncsu.edu/$\Tilde{}$jytzeng/software$\_$simreg.php}}); (ii) fastKM~\citep{marceau2015fast}, as implemented in the \texttt{FastKM} R package; and (iii) MAGEE~\citep{wang2020efficient}, as implemented in the  \texttt{MAGEE} R package.  MAGEE is the state-of-the-art scalable \GxE VC test with demonstrated superior performance compared to several set-based GxE methods.

In all simulations, 
we obtain the genotype design matrix $\mg\in\mathbb{R}^{n \times L}$ as follows. First, we employ the COSI software \citep{schaffner2005calibrating} to simulate $10{,}000$ haplotypes of SNP sequences mimicking the European population. We then form a SNP set of $L$ loci with minor allele frequency less than 1\% by randomly selecting $L$ SNPs without replacement. Finally, in each replicate, we generate the genotypes of $n$ individuals by randomly selecting two haploytpes with replacement. We consider $n=5{,}000$ and $L=100$ in the random effects simulations, and $n=20{,}000$ or  $100{,}000$ and $L=100$ or $400$ in the fixed effects simulations. We also consider a confounding factor $X\in \mathbb{R}^{n}$  and an environmental factor $E\in \mathbb{R}^{n}$, where each is generated from a standard normal distribution. Given $X$ and $E$, we then form the covariate design matrix $\mtx\in \mathbb{R}^{n\times 3}$ by column-combining the vector of ones, $X$, and $E$ together. 

\vspace{0.2cm}
\subsubsection{Random Effects Simulation Study}

Given the genotype design matrix $\mg$ of $n=5{,}000$ and $L=100$ loci and the $n\times P$ covariate design matrix $\mtx$ (where $P=3$), we simulate the outcome data $\vy$  according to the random effects model: 
$\vy = \mtx \vbeta + \mg \vb + \diag(E)\mg\vc +\ve$, %
where $\vbeta$ is set as the all ones vector of length $P$; $\vb$ is  generated from $\text{N}(\vzero, \tau\, \mi_{L})$;  $\ve$ is generated from $\text{N}(\vzero, \sigma\, \mi_{n})$; and $\sigma$ and $\tau$ are set to be 1. We set $\nu=0$ for Type I error analysis and $\nu>0$ for power analysis, where the actual value of $\nu$ is determined so that the empirical power is not too close to 1.  We simulate $N=1{,}000$ replicates and evaluate the results at the nominal level $\alpha=0.05$ for all analyses, except when assessing SEAGLE's Type I error rates at $\alpha=5\times 10^{-2}$, $5\times 10^{-3}$ and $5\times 10^{-4}$, 
where we consider $N=366{,}000$.

\begin{figure}[h]
\begin{center}
		\begin{minipage}{0.48\linewidth}
			\begin{center}
		\includegraphics[width=\textwidth]{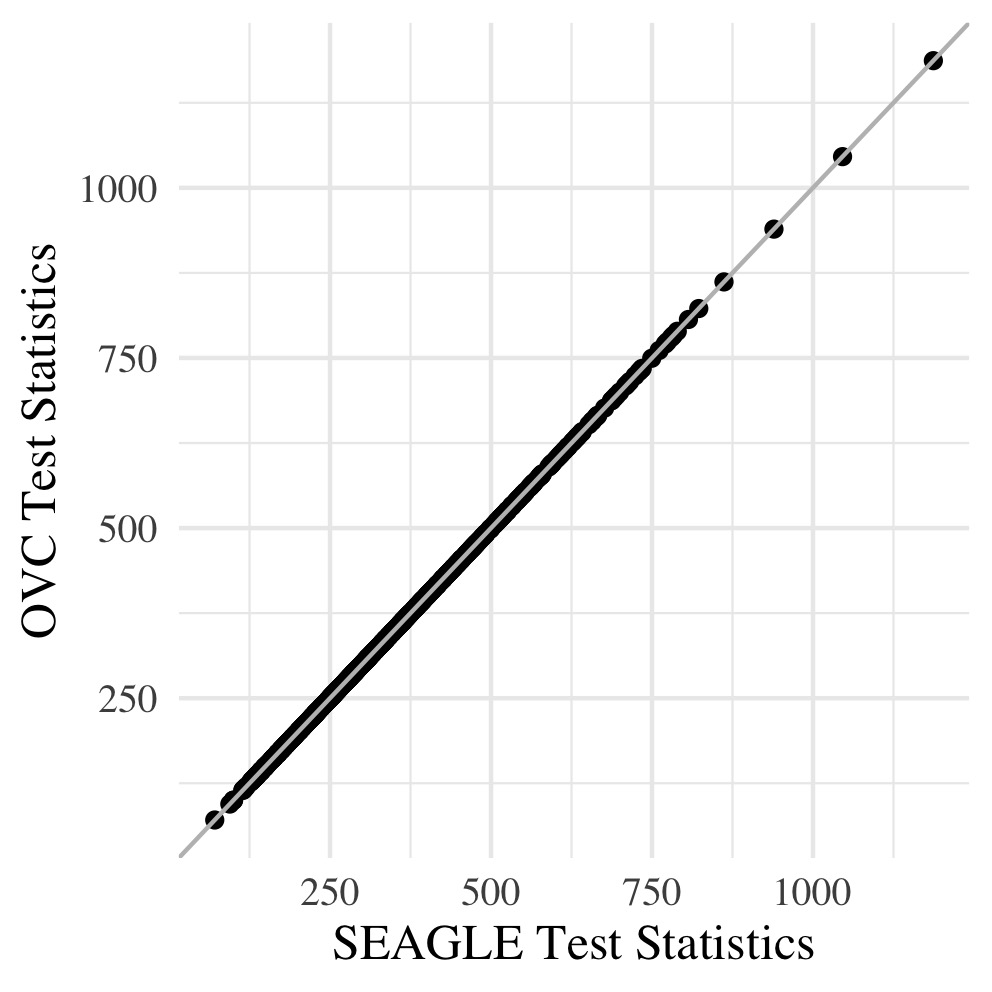}
			\end{center}
\begin{center}\textbf{(A)}\end{center}
		\end{minipage}
		\hspace{0.02\linewidth}
		\begin{minipage}{0.48\linewidth}
			\begin{center}
		\includegraphics[width=\textwidth]{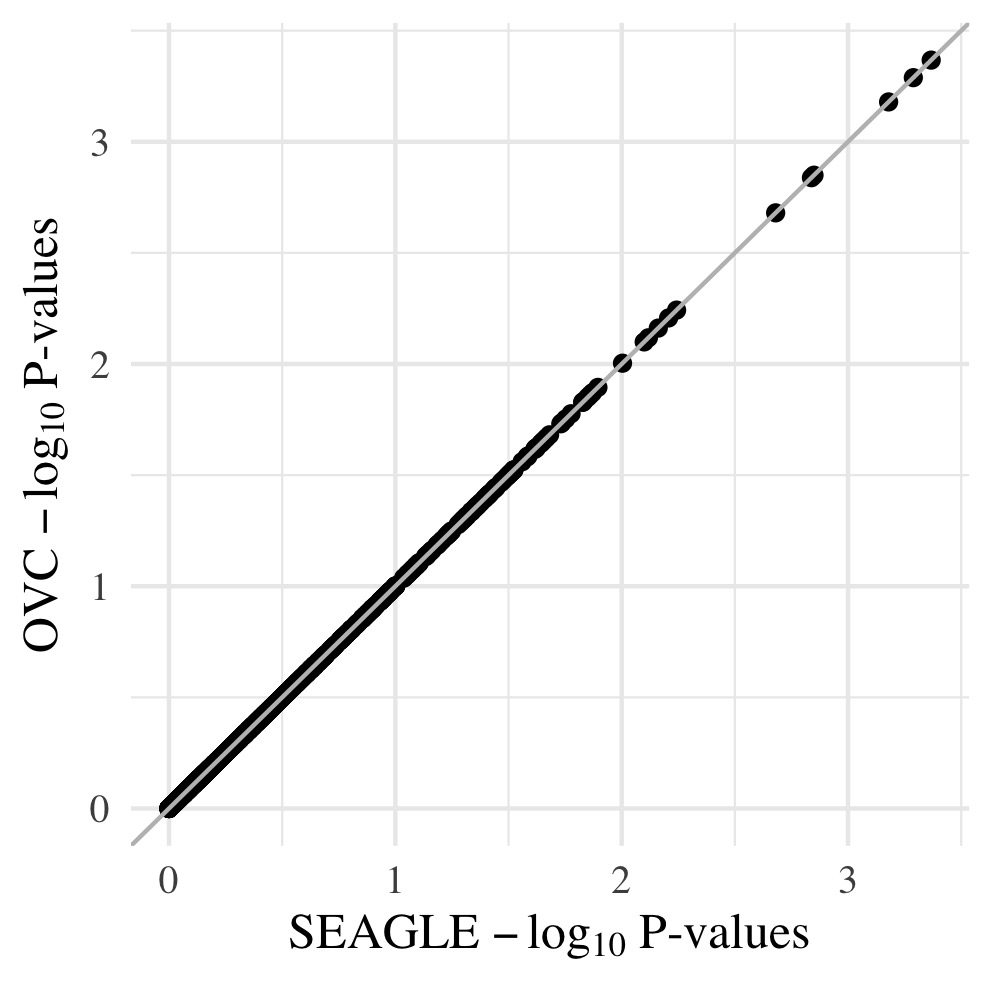}
			\end{center}
\begin{center}\textbf{(B)}\end{center}
		\end{minipage}
	\end{center}
\caption{\emph{SEAGLE vs. original \GxE VC (OVC) results based on $N=1{,}000$ replicates for $n=5{,}000$ observations and $L=100$ loci under $H_0:$ no \GxE effect ($\nu=0$).
Panels (A) and (B) show the scatter plots of the testing results computed from SEAGLE vs. those from OVC, depicting the ``exact" relationship between SEAGLE and OVC; panel (A) for test statistics $T$ and panel (B) for the p-values.}}
\label{fig:SEAGLEvsOVC}
\end{figure}

We start by examining the Type 1 error rate for SEAGLE 
with $N=366{,}000$ replicates.
Table \ref{table:sim1a_type1error_oursonly2} shows that SEAGLE provides reasonable control over the Type 1 error rate at varying $\alpha$-levels. 
%
%
%
Next, under $H_0:\nu=0$ and $\tau=\sigma=1$ with $N=1{,}000$ replicates, 
we compare the testing results of SEAGLE with OVC. 
Table \ref{tab:bias_mse_VC} shows the bias and the mean square error (MSE) of the estimated values for $\tau$ and $\sigma$  obtained from the SEAGLE and OVC REML EM algorithms. Both algorithms produce very small bias and MSE for $\tau$ and $\sigma$. Figures \ref{fig:SEAGLEvsOVC} (A) and (B) depict scatter plots of the score-like test statistics and p-values, respectively, produced by SEAGLE and OVC.  The figures show that SEAGLE and OVC produce identical test statistics and p-values, hence the ``exactness" of the SEAGLE algorithm.  

Since the data are generated from a random effects model under $H_0:\nu=0$, we can compute the ``true" score-like test statistic $T$ by evaluating \eqref{eqn:scoretest} at the true $\tau$ and $\sigma$ values, and obtaining the corresponding p-value. We refer to this as the ``Truth" and include it as a baseline approach. Supplementary Figure \ref{fig:sim1a_qq} depicts quantile-quantile plots (QQ plots) of the p-values obtained over $N=1{,}000$ replicates for Truth, SEAGLE, OVC, FastKM, and MAGEE.  All methods exhibit similar p-value behavior, and the red line for SEAGLE is very close to the light blue (OVC) and yellow lines (Truth).
In Table~\ref{tab:mse_pv}, we also compute the MSE 
of the p-values obtained from SEAGLE, OVC, FastKM, and MAGEE, compared to the Truth p-values. We observe that MAGEE produces p-values with larger MSE than the other methods.  
Supplementary Figure~\ref{fig:mse_re_pv}  shows the corresponding absolute relative error of the p-values for each method, computed by first taking the absolute difference between  a method's p-value and the Truth p-value, then dividing it by the Truth p-value. The boxplots 
suggest that 
MAGEE exhibits higher bias and greater variance than SEAGLE, OVC and FastKM. 

\begin{figure}[h]
	\begin{center}
		\begin{minipage}{.48\linewidth}
			\begin{center}
				\includegraphics[width=\textwidth]{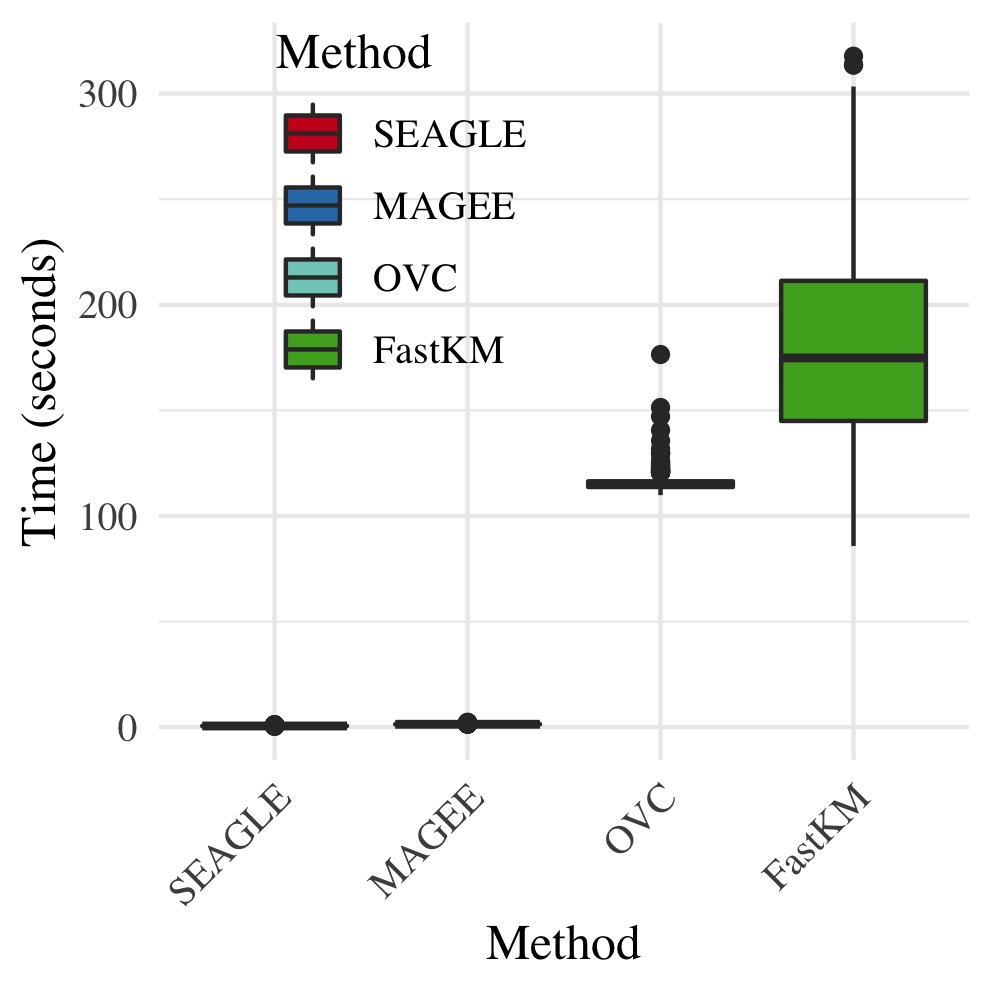}
			\end{center}
			\begin{center}(A)\end{center}
		\end{minipage}\hspace{0.01cm}
		\begin{minipage}{.48\linewidth}
			\begin{center}
				\includegraphics[width=\textwidth]{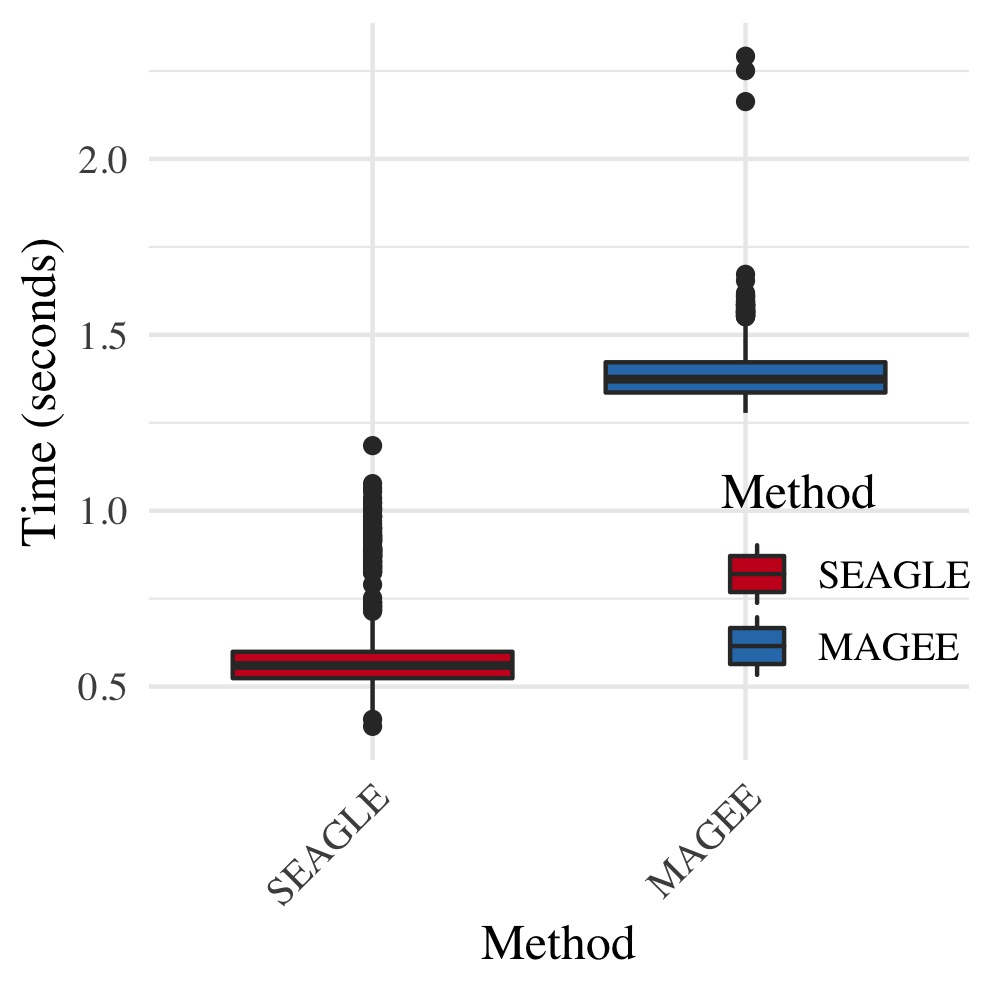}
			\end{center}
			\begin{center}(B)\end{center}
		\end{minipage}		
	\end{center}
	\caption{\emph{(A) Box plots of computation time in seconds to obtain a single p-value over $N=1{,}000$ replicates for $n=5{,}000$ observations and $L=100$ loci.  (B) Computation time in seconds to obtain a single p-value for SEAGLE and MAGEE.}} 
	\label{fig:sim1a_time}
\end{figure}

Regarding the computational cost, Figure \ref{fig:sim1a_time} (A) shows boxplots of the computation time in seconds required to obtain a single p-value for each of the methods over $N=1{,}000$ replicates with $\tau=\sigma=1$ and $\nu=0$. Figure \ref{fig:sim1a_time} (B) 
shows the same boxplots for just SEAGLE and MAGEE.  Results show that at $n=5{,}000$ observations and $L=100$ loci, SEAGLE is faster than MAGEE on average.
All replicates were computed on a 2013 Intel Core i5 laptop with a 2.70 GHz CPU and 16 GB RAM.  

\begin{figure}[h] 
	\begin{center}
		\begin{minipage}{.48\linewidth}
			\begin{center}
				\includegraphics[width=\textwidth]{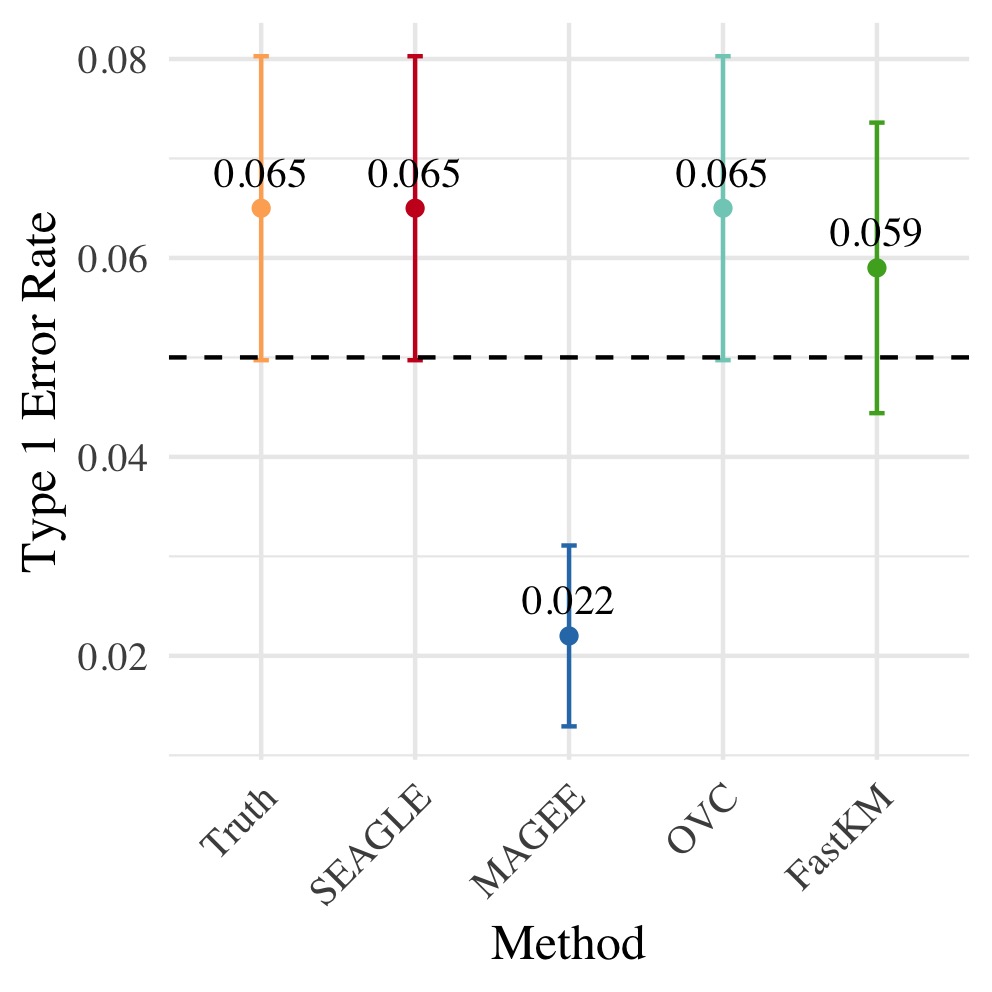}
			\end{center}
			\caption{\emph{Type 1 error at $\alpha = 0.05$ level for $N=1{,}000$ replicates with $n=5{,}000$ observations and $L=100$ loci with $\nu=0$.}}
			\label{fig:sim1a_type1error}
		\end{minipage} \hspace{0.05cm}
		\begin{minipage}{.48\linewidth}
			\begin{center}
				\includegraphics[width=\textwidth]{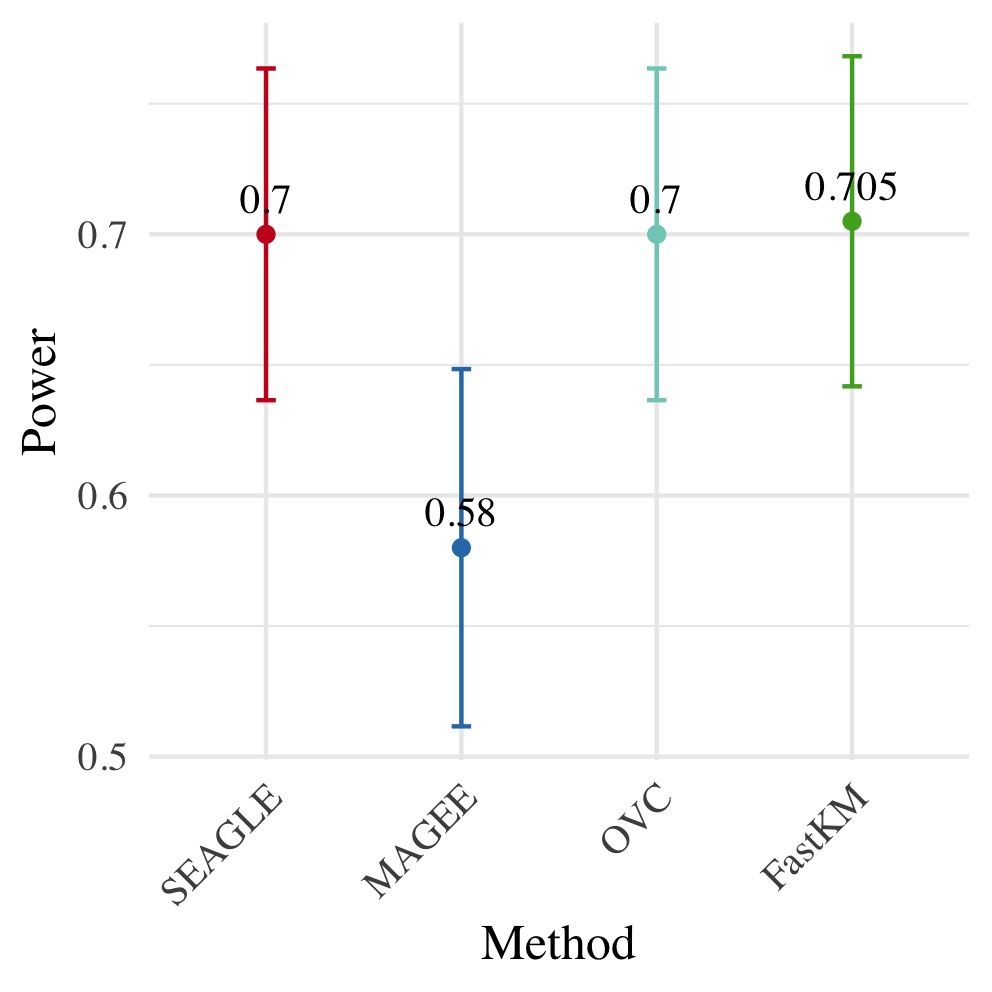}
			\end{center}
			\caption{\emph{Power at $\alpha = 0.05$ level over $N=200$ replicates with $n=5{,}000$ observations, $L=100$ loci, and $\nu=0.04$.}}
			\label{fig:sim1b_power}
		\end{minipage}
	\end{center}
\end{figure}
Figure \ref{fig:sim1a_type1error} shows the Type 1 error rate for each method under $H_0:\nu=0$.  SEAGLE performs identically to OVC with respect to Type 1 error rate at $\alpha=0.05$ while requiring only a fraction of the computation time, as demonstrated in Figure \ref{fig:sim1a_time}.  By contrast, MAGEE is nearly as fast as SEAGLE for $n=5{,}000$ and $L=100$ but produces much more conservative p-values at $\tau=\sigma=1$.

Figure \ref{fig:sim1b_power} shows the power for each method under the alternative hypothesis that $\nu > 0$.  SEAGLE again performs identically to OVC while requiring a fraction of the computation time.  By contrast, MAGEE is nearly as fast as SEAGLE for $n=5{,}000$ and $L=100$ but has lower power when $\tau=\sigma=1$ and $\nu=0.04$.

\vspace{0.2cm}
\subsubsection{Fixed Effects Simulation Study}

To study the performance of our proposed method when the data may not adhere to our model assumptions, we follow previous work \citep{marceau2015fast,wang2020efficient} and simulate data according to the fixed effects model with a given $\mtx$ and $\mg$:
$\vy = \mtx\vgamma_{\mtx} + \mg\vgamma_{G} + \diag(E)\mg\vgamma_{GE} + \ve$,
\noindent where $\vgamma_{\mtx}$ is the all ones vector of length $P = 3$, $\vgamma_{G} \in \mathbb{R}^{L}$, $\vgamma_{GE}\in \mathbb{R}^{L}$, and $\ve \sim \text{N}(\vzero, \sigma\, \mi_{n})$.  The entries of $\vgamma_{G}$ and $\vgamma_{GE}$ pertaining to causal loci are set to be $\gamma_{G}$ and $\gamma_{GE}$, respectively. The remaining entries of $\vgamma_{G}$ and $\vgamma_{GE}$ pertaining to non-causal loci are $0$.  
We consider $n=20{,}000$ or $100{,}000$ observations with $L=100$ or $400$.
We select the first $\ell$ loci to be causal (i.e., loci with both G and \GxE effects or just G effect).
We vary $\gamma_{G}$ over $0.5, 1$, and $1.5$ for the $\ell$ loci to study the impact of the G main effect sizes. We compare SEAGLE with MAGEE only since  OVC and fastKM are unable to work on the sample sizes considered here.

\begin{figure}[h]
	\begin{center}
		\includegraphics[width=\textwidth]{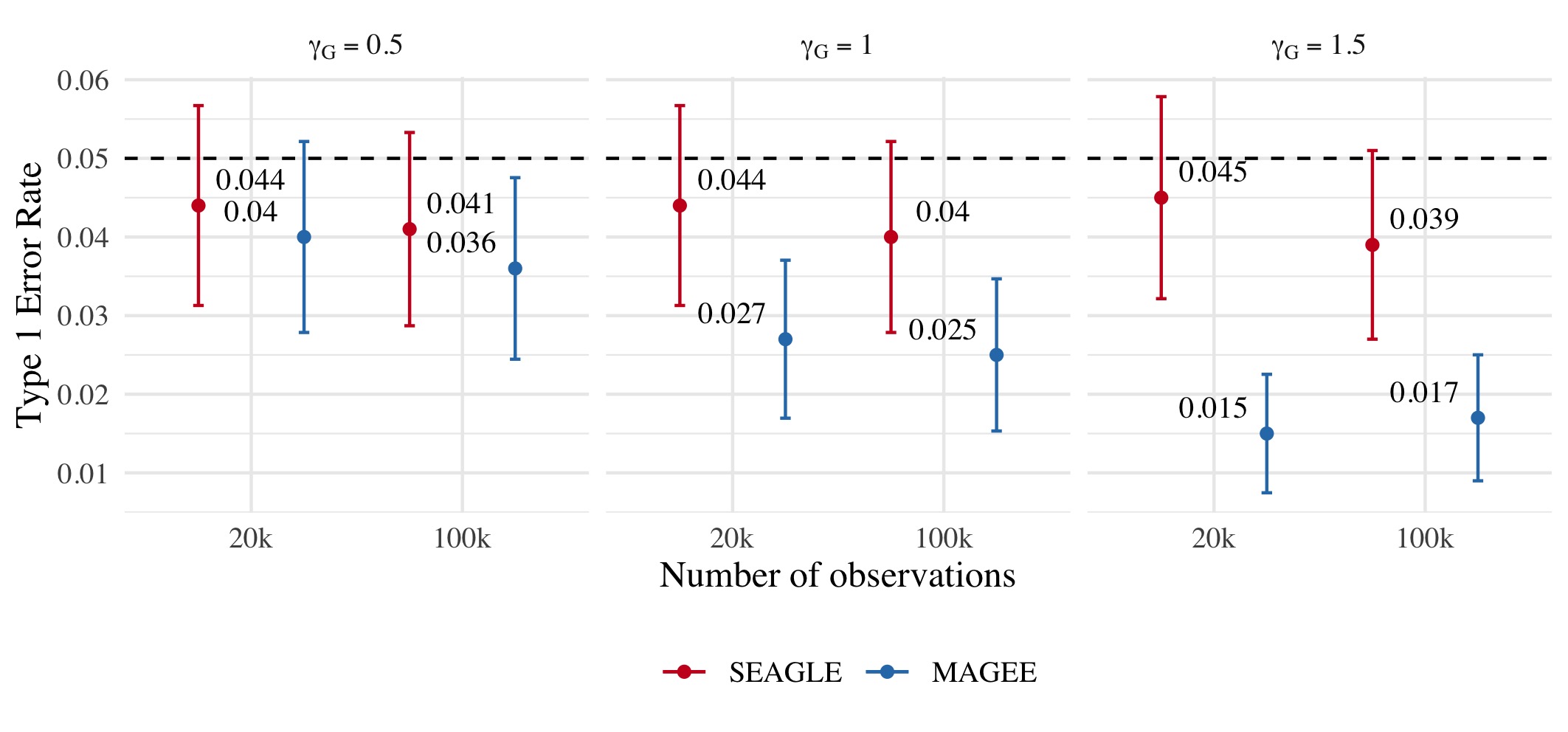}
	\end{center}
	\caption{\emph{Type 1 error at $\alpha=0.05$ for fixed effects simulations with $N=1{,}000$ replicates with $n=20{,}000$ and $n=100{,}000$ observations, and $L=100$ loci with $\gamma_{GE} = 0$ and varying values of $\gamma_{G}$.}}
	\label{fig:sim2a_type1error_0.05}
\end{figure}

We first evaluate the Type 1 error of SEAGLE by simulating $N=1{,}000$ replicates with $L=100$ for both $n=20{,}000$ and $100{,}000$, and setting $\gamma_{GE}=0$ for all loci while letting the first $\ell=40$ loci to have non-zero $\gamma_G$.
Figure \ref{fig:sim2a_type1error_0.05} depicts the Type 1 error rate at $\alpha = 0.05$ over varying values for $\gamma_{G}$.  While the Type 1 error rate for SEAGLE remains relatively unaffected by different $\gamma_{G}$ values, MAGEE produces more conservative p-values as $\gamma_{G}$ increases.  This is consistent with the MAGEE assumption requiring a small G main effect \citep{wang2020efficient}.  
Supplementary Figure \ref{fig:sim2a_qq} shows the corresponding quantile-quantile plots for the p-values obtained from SEAGLE and MAGEE.

\begin{figure}[h]
	\begin{center}
		\includegraphics[width=\textwidth]{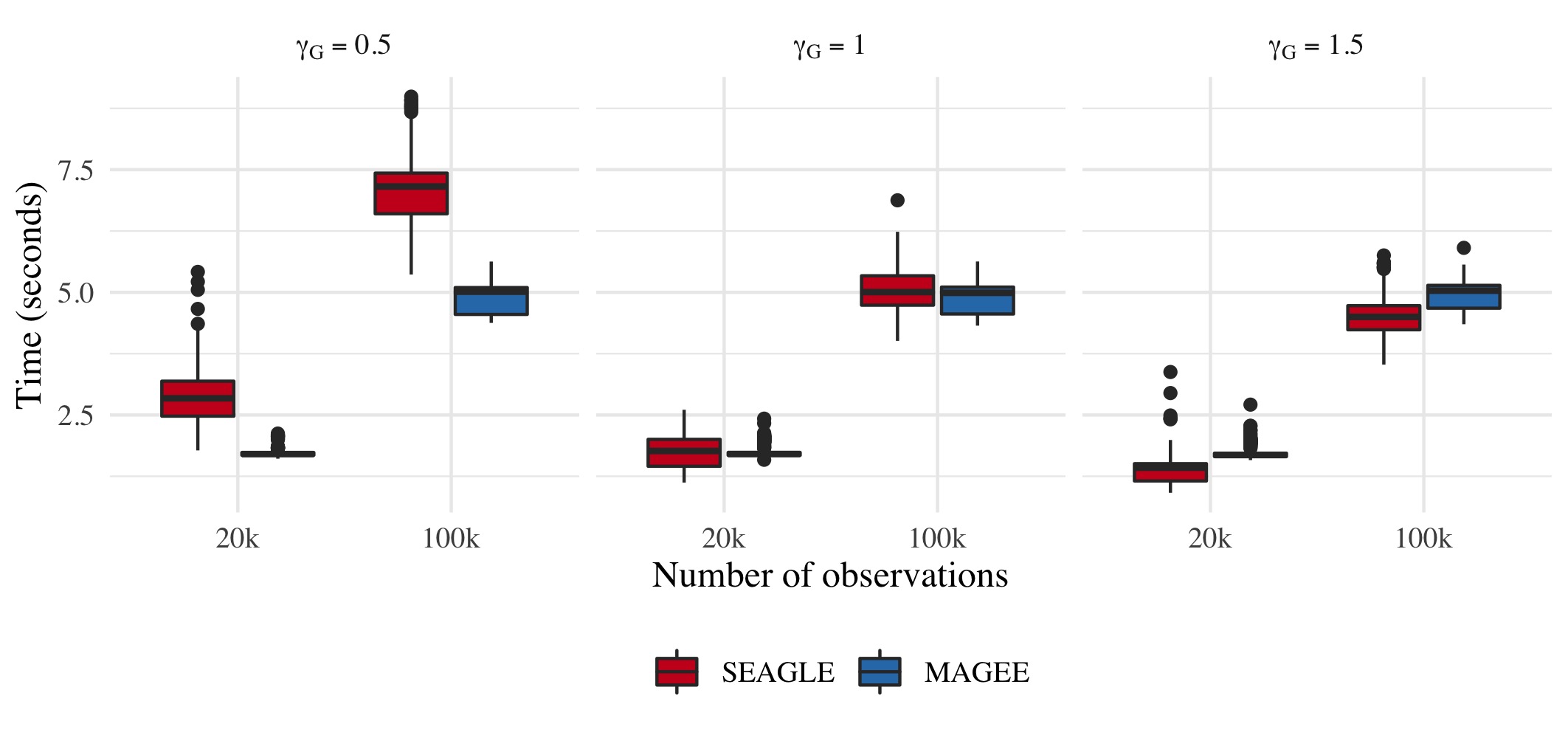}
	\end{center}
	\caption{\emph{Computation time in seconds for fixed effects simulations with $N=1{,}000$ replicates with $n=20{,}000$ observations and $L=100$ loci.}}  
	\label{fig:sim2a_time}
\end{figure}

Figure \ref{fig:sim2a_time} depicts boxplots of the computation time in seconds required to obtain a single p-value over the $N=1{,}000$ replicates for $n=20{,}000$ and $n=100{,}000$, and $L=100$, over varying values for $\gamma_{G}$.  All replicates were computed on a 2013 Intel Core i5 laptop with a 2.70 GHz CPU and 16 GB RAM.  
For $n=20{,}000$, SEAGLE is faster than MAGEE at larger values of $\gamma_{G}$ even though MAGEE computes an approximation to the test statistic $T$ and bypasses the traditional REML EM algorithm.  At smaller values of $\gamma_{G}$, however, SEAGLE requires a few seconds more than MAGEE. This is because smaller $\gamma_{G}$ values result in smaller $\tau$, and the REML EM algorithm converges slowly for $\tau$ values close to $0$. Supplementary  Figure \ref{fig:sim2a_varcomps} illustrates this empirically for $n=20{,}000$ with the estimated values of $\tau$ produced by the REML EM algorithm at different $\gamma_{G}$ values. 
%
These trends persist for $n=100{,}000$ observations.

\begin{figure}[h]
	\begin{center}
	\begin{minipage}{.48\linewidth}
	\begin{center}
		\includegraphics[width=\textwidth]{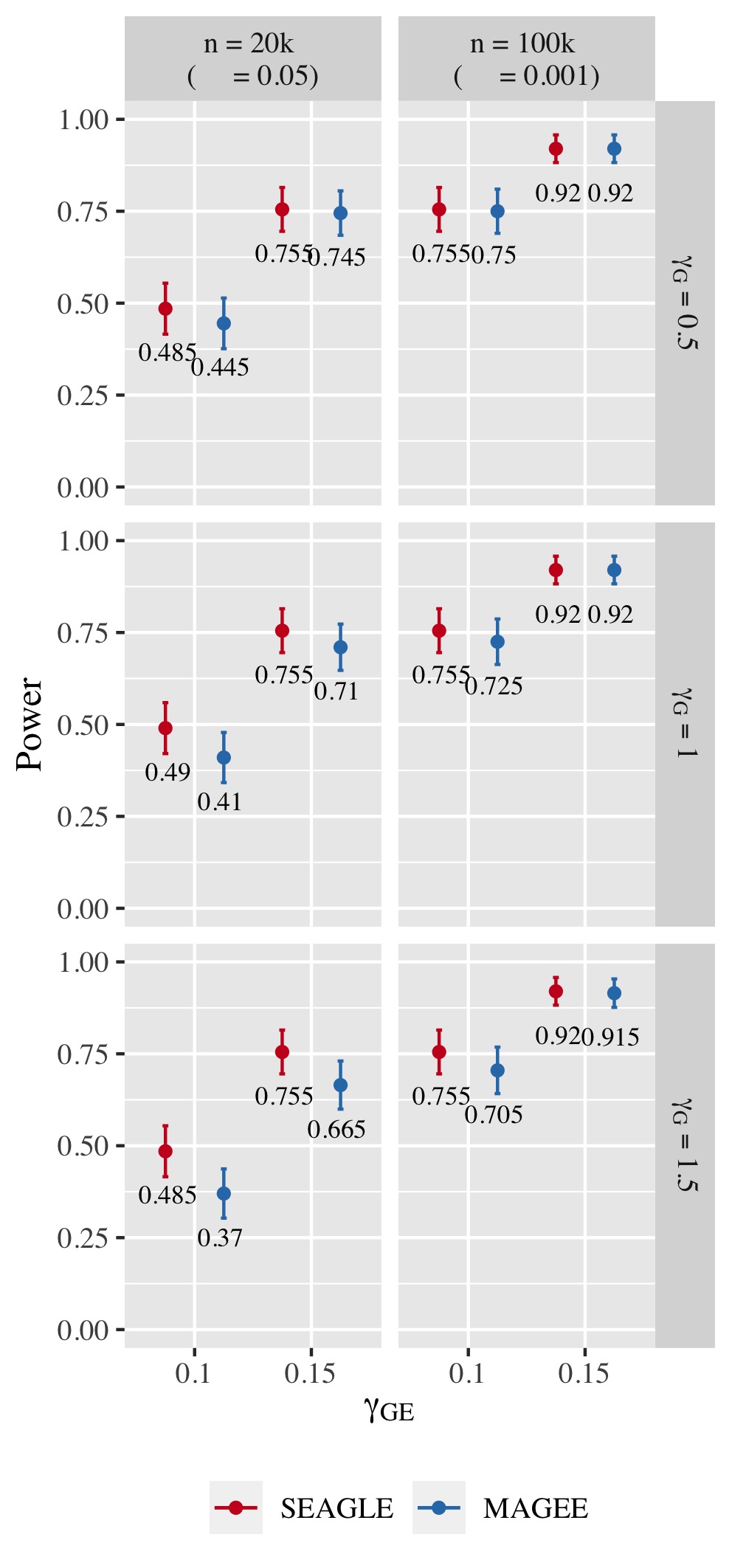}
		\vspace*{-96.3ex}
		\begin{center}
			$\alpha$ \quad\quad\quad\quad\quad\,\;\;\;\; $\alpha$ \quad\quad\quad
		\end{center}
		\vspace*{86ex}
	\end{center}
	\caption{\emph{Power at $\alpha=0.05$ and $\alpha=0.001$ for p-values from fixed genetic effects model over $N=200$ replicates with $n=20{,}000$ and $n=100{,}000$ observations, respectively, and $L=100$ loci.}}
	\label{fig:sim2b_power_L100}
	\end{minipage}\hspace{0.1cm}
	\begin{minipage}{.48\linewidth}
		\begin{center}
			\includegraphics[width=\textwidth]{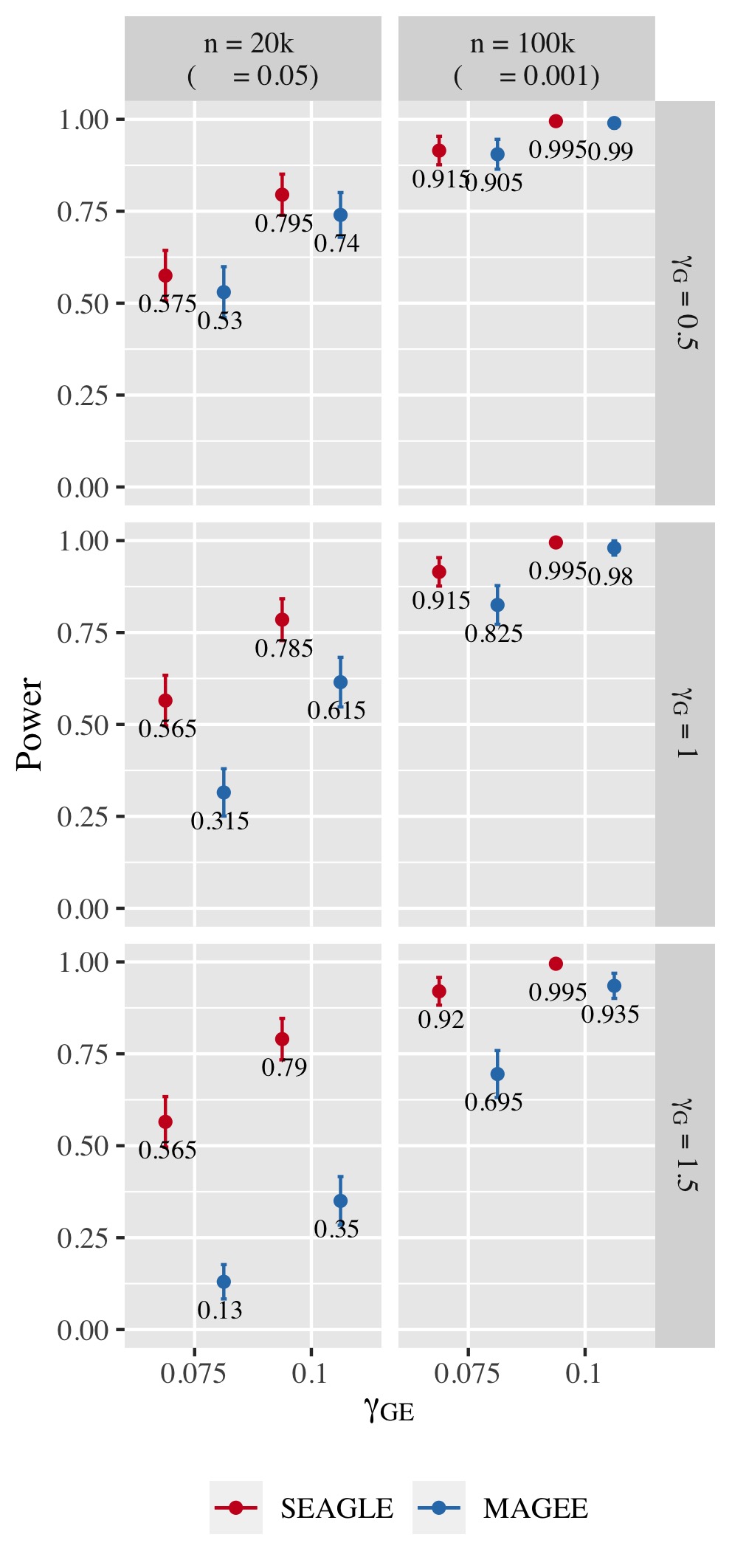}
			\vspace*{-96.3ex}
			\begin{center}
				$\alpha$ \quad\quad\quad\quad\quad\,\;\;\;\; $\alpha$ \quad\quad\quad
			\end{center}
			\vspace*{86ex}
		\end{center}
		\caption{\emph{Power at $\alpha=0.05$ and $\alpha=0.001$ for p-values from fixed genetic effects model  over $N=200$ replicates with $n=20{,}000$ and $n=100{,}000$ observations, respectively, and $L=400$ loci.}}
		\label{fig:sim2b_power_L400}
	\end{minipage}
	\end{center}
\end{figure}

For power evaluation, we simulate $N=200$ replicates with $L=100$ and $400$, and let
the first $\ell$ causal loci to have non-zero $\gamma_{G}$ and $\gamma_{GE}$. 
For $L=100$ loci, we set $\gamma_{GE}$ for the first $\ell=40$ causal loci to be $0.1$ or $0.15$.
For $L=400$ loci, we set $\gamma_{GE}$ for the first $\ell=120$ causal loci to be $0.075$ or $0.1$. These $\gamma_{GE}$ values are determined so that the power for $n=20{,}000$ at $\alpha=0.05$ is not close to 1. 
The values of $\gamma_{G}$ for the $\ell$ causal loci are set to be $0.5, 1.0$, and $1.5$ as before.

Figure \ref{fig:sim2b_power_L100} shows the power for $L=100$ loci.  
At $n=20{,}000$, SEAGLE exhibits better power than MAGEE at all combinations of $\gamma_{G}$ and $\gamma_{GE}$.  Moreover, the difference in power increases for larger values of $\gamma_{G}$ since MAGEE relies on the assumption that the G main effect size is small.  
At $n=100{,}000$ and the same values of $\gamma_{GE}$, we report the power at $\alpha=0.001$ instead of $0.05$ because the power at $\alpha=0.05$ is near 1 for both methods.
We see that both methods produce similar results although SEAGLE still outperforms MAGEE at slightly smaller values of $\gamma_{GE}$.  
%
Figure \ref{fig:sim2b_power_L400} shows the power for $L=400$ loci. Similar patterns of relative power performance are observed as in the case of $L=100$, except that the power difference between SEAGLE and MAGEE is more pronounced in $L=400$.

\vspace{0.2cm}
\subsection{Application to the Taiwan Biobank Data}
To illustrate the scalability of the \GxE VC test using SEAGLE, we apply SEAGLE and MAGEE to the Taiwan Biobank (TWB) data. 
TWB is a nationwide biobank project initiated in 2012 and has recruited more than 15,995 individuals. Peripheral blood specimens were extracted and genotyped using the Affymetrix Genomewide Axiom TWB array, which was designed specifically for a Taiwanese population.
%
We conduct the gene-based \GxE analysis and evaluate the interaction between gene and physical activity (PA) status on body mass index (BMI), adjusting for age, sex and the top 10 principal components for population substructure. The PA status is a binary indicator for with/without regular physical activity. Our \GxE analyses focuses on a subset of 11,664 unrelated individuals who have non-missing phenotype and covariate information. 
After PLINK quality control (i.e., removing SNPs with call rates $<0.95$ or Hardy-Weinberg Equilibrium p-value $<10^{-6}$), there are 589,867 SNPs remaining, which are mapped to genes according to the gene range list ``glist-hg19" 
from the PLINK Resources page at \url{https://www.cog-genomics.org/plink/1.9/resources}. There are a total of $13{,}260$ genes containing  $>$1 SNPs for G$\times$PA interaction analysis. 

\begin{figure}[ht]
	\begin{center}
		\includegraphics[width=0.7\textwidth]{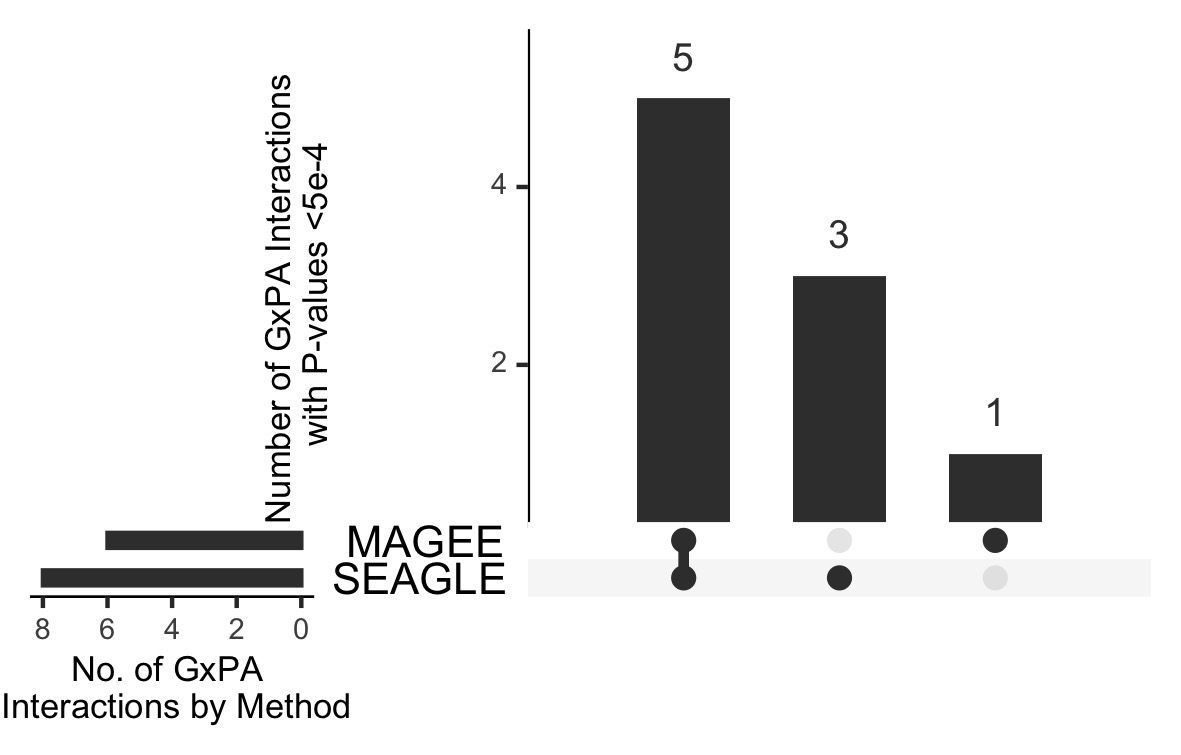}
	\end{center}
	\caption{\emph{Upset plot of significant G$\times$PA interactions identified by SEAGLE and MAGEE in the Taiwan Biobank at the $5\times 10^{-4}$ nominal level.}}
	\label{fig:twb_upset_bmi}
\end{figure}

The median run time of SEAGLE and MAGEE is 
$2.4$ and $1.3$ seconds, respectively, 
Both SEAGLE and MAGEE do not find any significant G$\times$PA interactions at the genome-wide Bonferroni threshold $0.05/13{,}260= 3.77\times 10^{-6}$. We hence discuss the results using a less stringent threshold, i.e., $5\times 10^{-4}$ and summarize the results in Figure \ref{fig:twb_upset_bmi} and Supplementary Table~\ref{table:twb_bmi}. SEAGLE and MAGEE identify 8 and 6 G$\times$PA interactions, respectively, among which 5 G$\times$PA results are identified by both methods (Figure \ref{fig:twb_upset_bmi}). The observation that SEAGLE identifies slightly more G$\times$PA effects than MAGEE generally agrees with the simulation findings. 
We use the GeneCards Human Gene Database (\url{www.genecards.org})~\citep{Stelzer2016} to explore the relevance of the identified genes with BMI or PA (see Supplementary Tables \ref{table:twb_bmi}).
Two of the 5 commonly identified genes, i.e., \textit{FCN2} and \textit{OCM}, have non-zero relevance scores (i.e., 0.56 and 0.91, respectively). 
For the 3 genes identified by SEAGLE only, i.e., 
\textit{ALOX5AP},
\textit{BCLAF1} and
\textit{PCDH17}, their relevance scores are 6.16, 0.26 and 1.54, respectively. The expression of \textit{ALOX5AP} has also been found associated with obesity and insulin resistance~\citep{kaaman2006alox5ap} as well as exercise-induced stress~\citep{hilberg2005transcription}. 
On the other hand, \textit{TBPL1} (identified by MAGEE only) is not in the GeneCards relevance list with BMI or PA.

\section{Discussion}
\label{sec:discussion}

We introduced SEAGLE, a scalable exact algorithm for performing set-based \GxE VC tests on large-scale biobank data. We achieve scalability and accuracy
by applying modern numerical analysis techniques, which include
avoiding the explicit formation of products and inverses  of large matrices. 
Our numerical experiments illustrate that SEAGLE produces Type 1 error rates and power that are identical to those of the original VC test \citep{tzeng2011studying}, 
while requiring a fraction of the computational cost.  Moreover, SEAGLE is well-equipped to handle the very large dimensions required for analysis of large-scale biobank data.  

State-of-the-art computational approaches such as \texttt{MAGEE} bypass the traditional time-consuming REML EM algorithm, and instead compute an approximation
to the score-like test statistic by assuming that the G main effect size is small. In practice, however, the G main effect size is often unknown. Our numerical experiments illustrate that SEAGLE generally achieves better Type 1 error and power with comparable computation time.


We highlight the fact that our timing experiments were performed
on a 2013 Intel Core i5 laptop with a 2.70 GHz CPU and 16 GB RAM.  Therefore,
SEAGLE performs efficient and exact set-based \GxE tests on biobank-scale data with $n=20{,}000$ and $n=100{,}000$ observations on ordinary laptops, without any need for high performance computational platforms.  
This makes SEAGLE broadly accessible to all researchers.  
Software for SEAGLE is publicly available as the \texttt{SEAGLE} 
package on GitHub (\url{https://github.com/jocelynchi/SEAGLE}), and will soon  be available
on the Comprehensive R Archive Network as well.

We conclude with a discussion of three avenues for future extensions.  First is the extension
of Model \eqref{eqn:mixedmodel} to generalized linear mixed models that can 
accommodate binary phenotype traits in 
addition to the continuous phenotype covered here. 

Second is the extension from a single environmental  factor
to a set of factors represented by $\me\in \mathbb{R}^{n\times q}$ with $q>1$.
 The corresponding extension of Model (\ref{eqn:mixedmodel}) is
\begin{eqnarray}
\vy&=&\mx \vbeta_X+\vh_E + \vh_G +\vh_{GE}+\veps, \text{with }\\\nonumber
&&\vh_E\sim N(0, \tau_E\mathbf{\Sigma}_{E}), \,
\vh_G\sim N(0, \tau_G\mathbf{\Sigma}_{G}),\,  
\vh_{GE}\sim N(0, \nu \mathbf{\Sigma}_{GE}),\,
\text{and } \veps\sim N(0, \sigma \mathbf{I}_{n}).
\label{eqn:RE_general}
\end{eqnarray}
Here $\mathbf{\Sigma}_E, \mathbf{\Sigma}_G, 
\mathbf{\Sigma}_{GE}\in\mathbb{R}^{n\times n}$ are variance matrices, where 
$\mathbf{\Sigma}_E = \me \me\Tra$, 
$\mathbf{\Sigma}_G = \mg \mg\Tra$ and 
$\mathbf{\Sigma}_{GE}$ is the element-wise products of $\mathbf{\Sigma}_G$ and $\mathbf{\Sigma}_E$. 
A straightforward adaptation of
SEAGLE's scalable REML EM algorithm to the EM algorithm in~\cite{wang2014complete} takes care of estimating the nuisance VC parameters.
Numerical analysis techniques analogous to the ones presented here
will be the foundation for the efficient extension to multi-E factors.

Third is the extension to other types of kernels~\citep{wang2014complete, broadaway2015kernel_PMID:25885490} of the current random effects framework,
which can be viewed as a special case of kernel machine regression with linear kernels.
As in~\cite{lumley2018fastskat} and \cite{wu2018scalable_PMID:29950019}, 
we will explore the potential of randomized numerical linear algebra,
by drawing on the authors' long standing expertise in 
the development of numerically stable, accurate and efficient
randomized matrix algorithms \citep{ChiI20b,DI18,ESSWAI10,HoIps14,HoIS17,IpsW12,SAI16,WIps14}.


%

\bibliographystyle{frontiersinHLTH_FPHY} 
\bibliography{references-GxE}

\begin{thebibliography}{34}
\expandafter\ifx\csname natexlab\endcsname\relax\def\natexlab#1{#1}\fi
\expandafter\ifx\csname urlstyle\endcsname\relax
  \expandafter\ifx\csname doi\endcsname\relax
  \def\doi#1{doi:\discretionary{}{}{}#1}\fi \else
  \expandafter\ifx\csname doi\endcsname\relax
  \def\doi{doi:\discretionary{}{}{}\begingroup \urlstyle{rm}\Url}\fi \fi
\expandafter\ifx\csname selectlanguage\endcsname\relax
  \def\selectlanguage#1{}\fi

\bibitem[{Ottman(1996)}]{ottman1996gene}
Ottman R.
\newblock Gene--environment interaction: definitions and study design.
\newblock {\em Preventive medicine\/} {\bf 25} (1996) 764--770.

\bibitem[{Hunter(2005)}]{hunter2005gene}
Hunter DJ.
\newblock Gene--environment interactions in human diseases.
\newblock {\em Nature reviews genetics\/} {\bf 6} (2005) 287--298.

\bibitem[{McAllister et~al.(2017)McAllister, Mechanic, Amos, Aschard, Blair,
  Chatterjee et~al.}]{mcallister2017current}
McAllister K, Mechanic LE, Amos C, Aschard H, Blair IA, Chatterjee N, et~al.
\newblock Current challenges and new opportunities for gene-environment
  interaction studies of complex diseases.
\newblock {\em American journal of epidemiology\/} {\bf 186} (2017) 753--761.

\bibitem[{Sulc et~al.(2020)Sulc, Mounier, G{\"u}nther, Winkler, Wood, Frayling
  et~al.}]{sulc2020quantification}
Sulc J, Mounier N, G{\"u}nther F, Winkler T, Wood AR, Frayling TM, et~al.
\newblock Quantification of the overall contribution of gene-environment
  interaction for obesity-related traits.
\newblock {\em Nature communications\/} {\bf 11} (2020) 1--13.

\bibitem[{Fav{\'e} et~al.(2018)Fav{\'e}, Lamaze, Soave, Hodgkinson, Gauvin,
  Bruat et~al.}]{fave2018gene}
Fav{\'e} MJ, Lamaze FC, Soave D, Hodgkinson A, Gauvin H, Bruat V, et~al.
\newblock Gene-by-environment interactions in urban populations modulate risk
  phenotypes.
\newblock {\em Nature communications\/} {\bf 9} (2018) 1--12.

\bibitem[{Ritz et~al.(2017)Ritz, Chatterjee, Garcia-Closas, Gauderman, Pierce,
  Kraft et~al.}]{ritz2017lessons_PMID:28978190}
Ritz BR, Chatterjee N, Garcia-Closas M, Gauderman WJ, Pierce BL, Kraft P,
  et~al.
\newblock Lessons learned from past gene-environment interaction successes.
\newblock {\em American journal of epidemiology\/} {\bf 186} (2017) 778--786.

\bibitem[{Lin et~al.(2013)Lin, Lee, Christiani, and Lin}]{lin2013testGESAT}
Lin X, Lee S, Christiani DC, Lin X.
\newblock Test for interactions between a genetic marker set and environment in
  generalized linear models.
\newblock {\em Biostatistics\/} {\bf 14} (2013) 667--681.

\bibitem[{Su et~al.(2017)Su, Di, and Hsu}]{su2017_PMID:27474101}
Su YR, Di CZ, Hsu L.
\newblock A unified powerful set-based test for sequencing data analysis of gxe
  interactions.
\newblock {\em Biostatistics\/} {\bf 18} (2017) 119--131.

\bibitem[{Lin et~al.(2016)Lin, Lee, Wu, Wang, Chen, Li et~al.}]{lin2016test}
Lin X, Lee S, Wu MC, Wang C, Chen H, Li Z, et~al.
\newblock Test for rare variants by environment interactions in sequencing
  association studies.
\newblock {\em Biometrics\/} {\bf 72} (2016) 156--164.

\bibitem[{Tzeng et~al.(2011)Tzeng, Zhang, Pongpanich, Smith, McCarthy, Sale
  et~al.}]{tzeng2011studying}
Tzeng JY, Zhang D, Pongpanich M, Smith C, McCarthy MI, Sale MM, et~al.
\newblock Studying gene and gene-environment effects of uncommon and common
  variants on continuous traits: a marker-set approach using gene-trait
  similarity regression.
\newblock {\em The American Journal of Human Genetics\/} {\bf 89} (2011)
  277--288.

\bibitem[{Zhao et~al.(2015)Zhao, Marceau, Zhang, and Tzeng}]{zhao2015assessing}
Zhao G, Marceau R, Zhang D, Tzeng JY.
\newblock Assessing gene-environment interactions for common and rare variants
  with binary traits using gene-trait similarity regression.
\newblock {\em Genetics\/} {\bf 199} (2015) 695--710.

\bibitem[{Wang et~al.(2015{\natexlab{a}})Wang, Maity, Luo, Neely, and
  Tzeng}]{wang2015_PMID:25538034}
Wang Z, Maity A, Luo Y, Neely ML, Tzeng JY.
\newblock Complete effect-profile assessment in association studies with
  multiple genetic and multiple environmental factors.
\newblock {\em Genetic epidemiology\/} {\bf 39} (2015{\natexlab{a}}) 122--133.

\bibitem[{Marceau et~al.(2015)Marceau, Lu, Holloway, Sale, Worrall, Williams
  et~al.}]{marceau2015fast}
Marceau R, Lu W, Holloway S, Sale MM, Worrall BB, Williams SR, et~al.
\newblock A fast multiple-kernel method with applications to detect
  gene-environment interaction.
\newblock {\em Genetic epidemiology\/} {\bf 39} (2015) 456--468.

\bibitem[{Wang et~al.(2020)Wang, Lim, Liu, Sung, Rao, Morrison
  et~al.}]{wang2020efficient}
Wang X, Lim E, Liu CT, Sung YJ, Rao DC, Morrison AC, et~al.
\newblock Efficient gene-environment interaction tests for large biobank-scale
  sequencing studies.
\newblock {\em Genetic Epidemiology\/} {\bf 44} (2020) 908--923.

\bibitem[{Liu et~al.(2009)Liu, Tang, and Zhang}]{liu2009anew}
Liu H, Tang Y, Zhang HH.
\newblock A new chi-square approximation to the distribution of non-negative
  definite quadratic forms in non-central normal variables.
\newblock {\em Computational Statistics \& Data Analysis\/} {\bf 53} (2009)
  853--856.

\bibitem[{Davies(1980)}]{davis1980}
Davies RB.
\newblock Algorithm as 155: The distribution of a linear combination of $\chi$
  2 random variables.
\newblock {\em Applied Statistics\/}  (1980) 323--333.

\bibitem[{Higham(2002)}]{Higham2002}
Higham NJ.
\newblock {\em Accuracy and stability of numerical algorithms\/} (Society for
  Industrial and Applied Mathematics (SIAM), Philadelphia, PA), second edn.
  (2002).

\bibitem[{Golub and Van~Loan(2013)}]{golub2013matrix}
Golub GH, Van~Loan CF.
\newblock {\em Matrix Computations 4th Edition\/}, vol.~4 (The Johns Hopkins
  University Press) (2013).

\bibitem[{Wang et~al.(2015{\natexlab{b}})Wang, Maity, Luo, Neely, and
  Tzeng}]{wang2014complete}
Wang Z, Maity A, Luo Y, Neely ML, Tzeng JY.
\newblock Complete effect-profile assessment in association studies with
  multiple genetic and multiple environmental factors.
\newblock {\em Genetic epidemiology\/} {\bf 39} (2015{\natexlab{b}}) 122--133.

\bibitem[{Schaffner et~al.(2005)Schaffner, Foo, Gabriel, Reich, Daly, and
  Altshuler}]{schaffner2005calibrating}
Schaffner SF, Foo C, Gabriel S, Reich D, Daly MJ, Altshuler D.
\newblock Calibrating a coalescent simulation of human genome sequence
  variation.
\newblock {\em Genome research\/} {\bf 15} (2005) 1576--1583.

\bibitem[{Stelzer et~al.(2016)Stelzer, Rosen, Plaschkes, Zimmerman, Twik,
  Fishilevich et~al.}]{Stelzer2016}
Stelzer G, Rosen N, Plaschkes I, Zimmerman S, Twik M, Fishilevich S, et~al.
\newblock The genecards suite: from gene data mining to disease genome sequence
  analyses.
\newblock {\em Current protocols in bioinformatics\/} {\bf 54} (2016) 1--30.

\bibitem[{Kaaman et~al.(2006)Kaaman, Ryd{\'e}n, Axelsson, Nordstr{\"o}m,
  Sicard, Bouloumie et~al.}]{kaaman2006alox5ap}
Kaaman M, Ryd{\'e}n M, Axelsson T, Nordstr{\"o}m E, Sicard A, Bouloumie A,
  et~al.
\newblock Alox5ap expression, but not gene haplotypes, is associated with
  obesity and insulin resistance.
\newblock {\em International journal of obesity\/} {\bf 30} (2006) 447--452.

\bibitem[{Hilberg et~al.(2005)Hilberg, Deigner, M{\"o}ller, Claus, Ruryk,
  Gl{\"a}ser et~al.}]{hilberg2005transcription}
Hilberg T, Deigner HP, M{\"o}ller E, Claus RA, Ruryk A, Gl{\"a}ser D, et~al.
\newblock Transcription in response to physical stress—clues to the molecular
  mechanisms of exercise-induced asthma.
\newblock {\em The FASEB journal\/} {\bf 19} (2005) 1492--1494.

\bibitem[{Broadaway et~al.(2015)Broadaway, Duncan, Conneely, Almli, Bradley,
  Ressler et~al.}]{broadaway2015kernel_PMID:25885490}
Broadaway KA, Duncan R, Conneely KN, Almli LM, Bradley B, Ressler KJ, et~al.
\newblock Kernel approach for modeling interaction effects in genetic
  association studies of complex quantitative traits.
\newblock {\em Genetic epidemiology\/} {\bf 39} (2015) 366--375.

\bibitem[{Lumley et~al.(2018)Lumley, Brody, Peloso, Morrison, and
  Rice}]{lumley2018fastskat}
Lumley T, Brody J, Peloso G, Morrison A, Rice K.
\newblock Fastskat: Sequence kernel association tests for very large sets of
  markers.
\newblock {\em Genetic epidemiology\/} {\bf 42} (2018) 516--527.

\bibitem[{Wu and Sankararaman(2018)}]{wu2018scalable_PMID:29950019}
Wu Y, Sankararaman S.
\newblock A scalable estimator of snp heritability for biobank-scale data.
\newblock {\em Bioinformatics\/} {\bf 34} (2018) i187--i194.

\bibitem[{Chi and Ipsen(2020)}]{ChiI20b}
Chi JT, Ipsen ICF.
\newblock A projector-based approach to quantifying total and excess
  uncertainties for sketched linear regression.
\newblock {\em submitted\/}  (2020).
\newblock ArXiv:1808.05924.

\bibitem[{Drineas and Ipsen(2019)}]{DI18}
Drineas P, Ipsen ICF.
\newblock Low-rank approximations do not need a singular value gap.
\newblock {\em SIAM J. Matrix Anal. Appl.\/} {\bf 40} (2019) 299--319.

\bibitem[{Eriksson-Bique et~al.(2011)Eriksson-Bique, Solbrig, Stefanelli,
  Warkentin, Abbey, and Ipsen}]{ESSWAI10}
Eriksson-Bique S, Solbrig M, Stefanelli M, Warkentin S, Abbey R, Ipsen I.
\newblock Importance sampling for a {Monte Carlo} matrix multiplication
  algorithm, with application to information retrieval.
\newblock {\em SIAM J. Sci. Comput.\/} {\bf 33} (2011) 1689--1706.

\bibitem[{Holodnak and Ipsen(2015)}]{HoIps14}
Holodnak JT, Ipsen ICF.
\newblock Randomized approximation of the {Gram} matrix: Exact computation and
  probabilistic bounds.
\newblock {\em SIAM J. Matrix Anal. Appl.\/} {\bf 36} (2015) 110--137.

\bibitem[{Holodnak et~al.(2018)Holodnak, Ipsen, and Smith}]{HoIS17}
Holodnak JT, Ipsen ICF, Smith RC.
\newblock A probabilistic subspace bound with application to active subspaces.
\newblock {\em SIAM J. Matrix Anal. Appl.\/} {\bf 39} (2018) 1208--1220.

\bibitem[{Ipsen and Wentworth(2014)}]{IpsW12}
Ipsen ICF, Wentworth T.
\newblock The effect of coherence on sampling from matrices with orthonormal
  columns, and preconditioned least squares problems.
\newblock {\em SIAM J. Matrix Anal. Appl.\/} {\bf 35} (2014) 1490--1520.

\bibitem[{Saibaba et~al.(2017)Saibaba, Alexanderian, and Ipsen}]{SAI16}
Saibaba AK, Alexanderian A, Ipsen ICF.
\newblock Randomized matrix-free trace and log-determinant estimators.
\newblock {\em Numer. Math.\/} {\bf 137} (2017) 353--395.

\bibitem[{Wentworth and Ipsen(2014)}]{WIps14}
Wentworth T, Ipsen ICF.
\newblock {Kappa$\_$SQ}: A {Matlab} package for randomized sampling of matrices
  with orthonormal columns.
\newblock {\em arxiv:1402:0642\/}  (2014).

\end{thebibliography}

\section*{Conflict of Interest Statement}
%
The authors declare that the research was conducted in the absence of any commercial or financial relationships that could be construed as a potential conflict of interest.
\section*{Author Contributions}
JTC, ICFI and JYT conceived the presented ideas and  study design. JTC implemented the methods and performed the numerical studies under the supervision of ICFI and JYT. JTC, THH, CHL, LSW, WPL, TPL and JYT contributed to data processing, analysis and result interpretation of the real data analysis. JTC, ICFI and JYT draft the manuscript with input from THH, CHL, LSW, WPL and TPL. All authors helped shape the research, discussed the results and contributed to the final manuscript.
%
%
\section*{Funding}
This work has been partially supported by National Science Foundation Grant DMS-1760374 (to JTC and ICFI), National Institutes of Health Grants U54 AG052427 (to LSW and WPL), U24 AG041689 (LSW, WPL and JYT), and P01 CA142538 (to JYT), and 
Taiwan Ministry of Science and Technology Grants MOST 106-2314-B-002-097-MY3 (to TPL) and MOST 109-2314-B-002-152 (to TPL).

\newpage


\begin{table}[ht]
	\caption{Type 1 error rate with $95\%$  confidence intervals (CIs) for SEAGLE over $N=366{,}000$ replicates for $n=5{,}000$ observations, $L=100$ loci, and variance components $\tau=\sigma=1$ and $\nu=0$.}
	\label{table:sim1a_type1error_oursonly2}
	\begin{center}
		\begin{tabular}{rrrc}
			\hline
			$\alpha$-Level & Type 1 Error & Std. Error & 95\%  CI \\ 
			\hline
			0.05 & 0.04784 & 0.00035 & (0.04715, 0.04853) \\ 
			0.005 & 0.00521 & 0.00012 & (0.00497, 0.00544) \\ 
			0.0005 & 0.00067 & 0.00004 & (0.00059, 0.00075) \\ 
			\hline
		\end{tabular}
	\end{center}
\end{table}

\begin{table}[ht]
	\caption{\emph{SEAGLE vs. original \GxE VC (OVC) results based on $N=1{,}000$ replicates for $n=5{,}000$ observations and $L=100$ loci under $H_0:$ no \GxE effect ($\nu=0$).  Table shows the bias and mean square error (MSE) of the estimated $\tau$ and $\sigma$ values, compared to the true values $\tau=\sigma=1$.}}
	\label{tab:bias_mse_VC}
	\begin{center}
		\begin{tabular}{llrr}
			\hline 
			&      & $\tau$ & $\sigma$ \\ \hline
			SEAGLE & Bias &   -3.06 $\times 10^{-4}$     &   -1.01 $\times 10^{-3}$      \\
			& MSE  &    4.37 $\times 10^{-2}$   &    4.18 $\times 10^{-4}$     \\
			OVC & Bias &    -6.93 $\times 10^{-4}$   &    -5.56 $\times 10^{-4}$      \\
			& MSE  &   4.36 $\times 10^{-2}$    &     1.05 $\times 10^{-4}$     \\ \hline
	\end{tabular}
	\end{center}
\end{table}

\begin{table}[ht]
	\caption{\emph{Mean squared error (MSE) of the p-values obtained from different GxE VC tests, compared to the ``Truth" p-values. Results are obtained with $\tau=\sigma=1$ under $H_0:\nu=0$ over $N=1{,}000$ replicates with $n=5{,}000$ observations and $L=100$ loci.}}
	\label{tab:mse_pv}
	\begin{center}
	\begin{tabular}{lr}\hline
		& MSE of P-value  \\ \hline
		SEAGLE &        $3.49\times 10^{-4}$          \\
		MAGEE & $224.96\times 10^{-4}$          \\
		OVC & $3.49\times 10^{-4}$          \\
		FastKM & $17.88\times 10^{-4}$          \\\hline
	\end{tabular}
	\end{center}
\end{table}

\end{document}